\documentclass[10pt]{article}
\usepackage{amsfonts,amsmath,amssymb}
\usepackage{float,braket}
\usepackage{pslatex}
\usepackage{graphicx,color}
\usepackage{subfigure}
\usepackage{graphicx,wrapfig}
\usepackage[parfill]{parskip}
\usepackage[small,bf]{caption2}
\usepackage{a4wide}

\newcommand{\lms}{\Lambda_{\overline{\mbox{\tiny{MS}}}}}
\newcommand{\gf}{\ensuremath{\mathrm{gf}}}
\newcommand{\YM}{\ensuremath{\mathrm{YM}}}
\newcommand{\e}{\ensuremath{\mathrm{e}}}
\newcommand{\s}{\ensuremath{\mathrm{s}}}
\newcommand{\p}{\partial}

\newcommand{\quadr}{\ensuremath{\mathrm{quad}}}

\newcommand{\AGZ}{\ensuremath{\mathrm{AGZ}}}

\newcommand{\CGZ}{\ensuremath{\mathrm{CGZ}}}
\newcommand{\GZ}{\ensuremath{\mathrm{GZ}}}

\newcommand{\phys}{\ensuremath{\mathrm{phys}}}
\newcommand{\vac}{\ensuremath{\mathrm{vac}}}
\newcommand{\Tr}{\ensuremath{\mathrm{Tr}}}

\newcommand{\ii}{\ensuremath{\mathrm{i}}}

\renewcommand{\d}{\ensuremath{\mathrm{d}}}

\newcommand{\MSbar}{\overline{\mbox{MS}}}
\definecolor{Rood}{rgb}{1, 0, 0} 

\begin{document}
\title{{\bf The dynamical origin of the refinement  of  the Gribov-Zwanziger theory}}
\author{D.~Dudal \thanks{david.dudal@ugent.be}\,\,$^a$, S.~P.~Sorella\thanks{sorella@uerj.br}\,\,$^b$, N.~Vandersickel \thanks{nele.vandersickel@ugent.be}\,\,$^a$\\
\\
\small $^a$ \textnormal{Ghent University, Department of Physics and Astronomy} \\
\small \textnormal{Krijgslaan 281-S9, 9000 Gent,Belgium}\\
\\
\small $^b$  \textnormal{Departamento de F\'{\i }sica Te\'{o}rica, Instituto de F\'{\i }sica, UERJ - Universidade do Estado do Rio de Janeiro}\\
\small   \textnormal{Rua S\~{a}o Francisco Xavier 524, 20550-013 Maracan\~{a}, Rio de Janeiro, Brasil }\normalsize}

\date{}
\maketitle

\abstract{In recent years, the Gribov-Zwanziger action was refined by taking into account certain  dimension 2 condensates. In this fashion, one succeeded in bringing the gluon and the ghost propagator obtained from the GZ model in qualitative and quantitative agreement with the lattice data. In this paper, we shall elaborate further on this aspect. First, we shall show that more dimension 2 condensates can be taken into account than considered so far and,  in addition, we shall give firm evidence that  these condensates are in fact present by discussing the effective potential. It follows thus that the Gribov-Zwanziger action dynamically transforms  itself into the refined version, thereby showing that the continuum nonperturbative Landau gauge fixing,  as implemented by the Gribov-Zwanziger approach, is consistent with lattice simulations. }

\section{Introduction}
The infrared behavior of  the gluon and ghost propagator has received a lot of interest in recent years, in particular in the Landau gauge. Many of the discussions were evolved around the zero momentum value of the gluon propagator and the infrared enhancement of the ghost. The common belief is now that in 4D and 3D the ghost propagator displays no enhanced behavior,  while  the gluon propagator exhibits positivity violation, being suppressed in the infrared.  Moreover, it attains a non-vanishing value at zero momentum. These results are supported by many lattice data \cite{Cucchieri:2007md,Bogolubsky:2007ud,Cucchieri:2007rg,Dudal:2010tf,Bogolubsky:2009dc,Cucchieri:2010xr,Bornyakov:2009ug,Maas:2008ri} as by many analytical approaches \cite{Aguilar:2004sw,Aguilar:2008xm,Fischer:2008uz,Binosi:2009qm,Boucaud:2008ky,Pene:2009iq,RodriguezQuintero:2011vp,Tissier:2010ts,Tissier:2011ey}. Such propagators  have been used to extract  results on the spectrum of gauge theories, see e.g.~\cite{Roberts:1994dr,Dudal:2010cd}. In particular, in the Gribov-Zwanziger (GZ) framework,  which accounts for the existence of (most of) the Gribov copies in the path integral \cite{Zwanziger:1989mf,Zwanziger:1992qr}, this behavior of the ghost and gluon propagator was explained by taking into account the existence of a certain BRST invariant dimension 2 condensate \cite{Dudal:2008sp,Dudal:2008rm}. This was called the refined Gribov-Zwanziger framework. This particular condensate was investigated as it corresponds to  a BRST invariant operator. However, one could go one step further. The Gribov-Zwanziger action has a softly broken BRST symmetry \cite{Zwanziger:1989mf,Dudal:2008sp}. Despite this, it is still renormalizable thanks to a wide set of Ward identities obeyed by the GZ action. Therefore, one could ask why one would only investigate $d=2$ BRST invariant condensates? \\
\\
In fact, there exists a whole range of $d=2$ condensates overlooked so far,  which might be taken into account.  In this paper, we shall firstly explore these condensates and show that they affect the gluon and the ghost propagator,  although not altering their qualitative behavior. The gluon propagator is still  suppressed and non-zero at zero momentum, and the ghost propagator is not enhanced. Secondly, , we shall also be able,  for the first time, to calculate the effective action with the help of the local composite operator (LCO) formalism at lowest order and give arguments that there is in fact condensation. We shall show that the minimum of the effective potential including the condensates is a non trivial minimum, i.e.~in this minimum the condensates are present, leading to a dynamical transformation of the GZ action into the refined GZ action.\\
\\
This paper is organized as follows. In section 2, we shall briefly review the construction of the Gribov-Zwanziger action. The first main point of this paper shall be proven in section 3, i.e.~there can be more $d=2$ condensates affecting the GZ action than considered so far. The second main point of this paper is presented in section 4,  namely: the construction of the effective action with the help of the local composite opeator (LCO) formalism \cite{Verschelde:2001ia,Verschelde:1995jj}. We first explain the LCO formalism and then apply it to the GZ action with the inclusion of  the set of $d=2$ condensates. We then show that searching for extrema of the effective action automatically leads to nonvanishing condensates, i.e.  to the refining of the GZ action. In section 5, we present the form of the gluon and the ghost propagator and show that  they are in qualitative agreement with the current lattice data, irrespective of the details of the condensation. In section 6 we collect our conclusion. Technical details are provided in a series of appendices.

\section{ Summary of the Gribov-Zwanziger formalism}
The Gribov-Zwanziger action takes into account the existence of Gribov copies by restricting the  domain  of integration  in the functional integral to the Gribov region $\Omega$, which is defined as the set of field configurations fulfilling the Landau gauge condition and for which the Faddeev-Popov operator,
\begin{equation}
\mathcal M^{ab} = -\p_\mu \left( \p_\mu  \delta^{ab} + g f_{abc} A_\mu^c\right)\;,
\end{equation}
is strictly positive. In \cite{Zwanziger1989} it has been firstly shown that this restriction to the Gribov region $\Omega$ can be established by considering the following (local) action
\begin{eqnarray}\label{GZaction}
S_\GZ &=& S_0 + S_\gamma
\end{eqnarray}
with
\begin{eqnarray}
S_{0} &=&S_{\mathrm{YM}}+S_{\gf}  +\int \d^d x \left( \overline{\varphi }_{\mu }^{ac} \partial _{\nu} D_\nu^{am}\varphi _{\mu
}^{mc} -\overline{\omega }_{\mu }^{ac} \partial _{\nu } D_\nu^{am} \omega _{\mu }^{mc}
  -g\left( \partial _{\nu }\overline{\omega }_{\mu}^{ac}\right) f^{abm}\left( D_{\nu }c\right) ^{b}\varphi _{\mu
}^{mc}\right) \nonumber \;, \\
S_{\gamma}&=& -\gamma ^{2}g\int\d^d x\left( f^{abc}A_{\mu }^{a}\varphi _{\mu }^{bc}+f^{abc}A_{\mu}^{a}\overline{\varphi }_{\mu }^{bc} + \frac{d}{g}\left(N^{2}-1\right) \gamma^{2} \right) \;.
\end{eqnarray}
with $S_{\YM}$ the classical Yang-Mills action and $S_\gf$ the Landau gauge fixing
\begin{eqnarray}
S_{\YM} &=& \frac{1}{4}\int \d^d x F^a_{\mu\nu} F^a_{\mu\nu}\;, \nonumber\\
S_{\gf} &=& \int \d^d x\,\left( b^a \p_\mu A_\mu^a +\overline c^a \p_\mu D_\mu^{ab} c^b \right)\,.
\end{eqnarray}
The fields $\left( \overline{\varphi }_{\mu}^{ac},\varphi_{\mu}^{ac}\right) $ are a pair of complex conjugate bosonic fields, while $\left( \overline{\omega }_{\mu}^{ac},\omega_{\mu}^{ac}\right) $ are anticommuting fields.  We recall that we can simplify the notation of the additional fields $\left( \overline \varphi_\mu^{ac},\varphi_\mu^{ac},\overline \omega_\mu^{ac},\omega_\mu^{ac}\right) $ in $S_0$ as $S_0$ displays a symmetry with respect to the composite index $i=\left( \mu,c\right)$. Therefore, we can set
\begin{equation}
\left( \overline \varphi_\mu^{ac},\varphi_\mu^{ac},\overline \omega_\mu^{ac},\omega_\mu^{ac}\right) =\left( \overline \varphi_i^a,\varphi_i^a,\overline \omega_i^a,\omega_i^a \right)\,,
\end{equation}
and thus
\begin{eqnarray}
S_{0}&=&S_\YM + S_\gf + \int \d^d x \left( \overline \varphi_i^a \p_\mu \left( D_\mu^{ab} \varphi^b_i \right)  - \overline \omega_i^a \p_\mu \left( D_\mu^{ab} \omega_i^b \right) - g f^{abc} \p_\mu \overline \omega_i^a     (D_\mu^{bd} c^d)  \varphi_i^c \right)  \;,
\end{eqnarray}
The BRST variations of all the fields are given by,
\begin{align}\label{BRST11}
sA_{\mu }^{a} &=-\left( D_{\mu }c\right) ^{a}\,, & sc^{a} &=\frac{1}{2}gf^{abc}c^{b}c^{c}\,,   \nonumber \\
s\overline{c}^{a} &=b^{a}\,,&   sb^{a}&=0\,,  \nonumber \\
s\varphi _{i}^{a} &=\omega _{i}^{a}\,,&s\omega _{i}^{a}&=0\,,\nonumber \\
s\overline{\omega}_{i}^{a} &=\overline{\varphi }_{i}^{a}\,,& s \overline{\varphi }_{i}^{a}&=0  \,.
\end{align}
 The massive parameter $\gamma$, called the Gribov parameter, is not an independent parameter of the theory, being determined in a self-consistent way by
the following gap equation, commonly known as the horizon condition,
\begin{eqnarray}\label{horizonconditon2}
     \braket{g f^{abc} A^a_{\mu} \varphi^{bc}_{\mu}} + \braket{g f^{abc} A^a_{\mu} \overline{\varphi}^{bc}_{\mu}} + 2 \gamma^2 d (N^2 -1)   = 0
\end{eqnarray}
 which ensures the restriction to the Gribov region. This gap equation can also be written as
\begin{eqnarray}\label{gapgamma}
\frac{\p \Gamma}{\p \gamma^2} &=& 0\;,
\end{eqnarray}
with $\Gamma$ the quantum action defined as
\begin{eqnarray}
\e^{-\Gamma} &=& \int [\d\Phi] \e^{-S_\GZ} \;,
\end{eqnarray}
where $\int [\d\Phi]$ stands for the integration over all the fields. The action $S_\GZ$ is renormalizable. For the benefit of the reader, we have  presented the full algebraic  proof of the renormalization of this action in the Appendix A, since we have to built on this anyway later on. Let us also mention that,  recently, an alternative approach was worked out to study the renormalizability of the GZ action \cite{Capri:2010hb,Capri:2011wp}. In this paper, we shall however follow the original approach of e.g.~\cite{Dudal:2010fq}.
\\
\\
We recall that the GZ action breaks the BRST symmetry explicitly \cite{Zwanziger:1989mf,Dudal:2008sp}. This is due to the $\gamma$-dependent term, $S_\gamma$, and one can easily check from  \eqref{BRST11} and \eqref{GZaction} that,
\begin{equation}\label{breaking}
s S_\GZ = s (S_{0} +  S_{\gamma}) ~=~ s (  S_{\gamma}) ~=~- g \gamma^2 \int \d^d x f^{abc} \left( A^a_{\mu} \omega^{bc}_\mu -
 \left(D_{\mu}^{am} c^m\right)\left( \overline{\varphi}^{bc}_\mu + \varphi^{bc}_{\mu}\right)  \right)\,.
\end{equation}

\section{Further refining  of the Gribov-Zwanziger action}

\subsection{Introduction}
So far,  the GZ action has been refined \cite{Dudal:2008sp} by investigating  the BRST invariant $d=2$ condensate $\Braket{\overline{\varphi}^a_i \varphi^a_{i} - \overline{\omega}^a_i \omega^a_i}$ and the well known condensate $\braket{A_\mu^a A_\mu^a}$. The first condensate  assures  that the gluon propagator is non-zero at zero momentum \cite{Dudal:2008sp}, while the second condensate is indispensable  in order to find a good quantitative agreement with the lattice data, see \cite{Dudal:2010tf,tobewritten}. The  resulting action, called the Refined Gribov-Zwanziger action (RGZ), gives rise to a ghost propagator which behaves like $1/p^2$ for small $p^2$,  and to the tree level gluon propagator given by
\begin{eqnarray}\label{gluonprop2}
  \; \Braket{ A^a_{\mu}(p) A^b_{\nu}(-p)} &=& \frac{1}{p^2+ m^2 + \frac{2g^2 N \gamma^4}{p^2 + M^2}}\left[\delta_{\mu\nu} - \frac{p_{\mu}p_{\nu}}{p^2} \right]\delta^{ab} \nonumber\\
 &=& \underbrace{\frac{p^2 + M^2}{p^4 + (M^2+m^2)p^2 + 2 g^2 N \gamma^4 + M^2 m^2 }}_{\mathcal{D}(p^2)}\left[\delta_{\mu\nu} - \frac{p_{\mu}p_{\nu}}{p^2} \right]\delta^{ab} \;.
\end{eqnarray}
whereby $M^2$ is the mass related to the condensate $\Braket{\overline{\varphi}^a_i \varphi^a_{i} - \overline{\omega}^a_i \omega^a_i}$ and $m^2$ to $\braket{A_\mu^a A_\mu^a}$. We clearly observe that this propagator is non-vanishing at zero momentum due to the presence of the mass $M^2$.\\
\\
However, as the GZ action breaks the BRST symmetry anyhow, see expression \eqref{breaking}, there is a priori no need to keep the  operators $\overline{\varphi}^a_i \varphi^a_{i}$ and  $\overline{\omega}^a_i \omega^a_i$ in a BRST invariant combination, i.e.  $\left( \overline{\varphi}^a_i \varphi^a_{i} - \overline{\omega}^a_i \omega^a_i \right)= s\left(\overline{\omega}^a_i \varphi^a_{i} \right)$. In fact, we can split the operator into two separate operators,  coupled to different sources. Moreover, there are also other $d=2$ operators, which were overlooked so far. In fact, all possible renormalizable $d=2$ operators $\mathcal O_i$ in the GZ action, which have ghost number zero, are given by\footnote{We are not considering the operator $\overline c^a c^a$ here. A $\braket{\overline c^a c^a}$ condensate would result in massive ghosts, something which is clearly excluded by lattice simulations. If $\overline c^a c^a$ is not directly coupled to the theory, it can neither radiatively appear due to a shift symmetry of the underlying action, viz.~$\overline c^a \to \overline c^a + cte$, with $cte$ a constant Grassmann parameter. }
\begin{equation}\label{operators}
\mathcal O_i = \{  A_\mu A_\mu,  \varphi_i^a  \varphi_i^a, \varphi_i^a \overline \varphi_i^a,   \overline{\varphi}^a_i \overline \varphi^a_i , \overline \omega^a_i \omega^a_i \}\;.
\end{equation}
We shall only investigate condensates which are fully contracted over the indices  $(a,i)$, e.g.~like $\varphi_i^a \overline \varphi_i^a = \varphi_\mu^{ac} \overline \varphi_\mu^{ac}$.   However, it is possible to make different contractions over the color indices as is shown in \cite{Gracey:2010cg}. Therefore, if one wants to be absolutely complete, one would have to take into account all possible color contractions. Unfortunately, this would be hopelessly complicated.   Though, we hope that a good description of the IR behavior of the gluon and ghost propagator has been captured by taking into account only one color combination.  Comparison with lattice data in 3D and 4D seems to confirm this, at least so far, \cite{Dudal:2010tf,tobewritten}.\\\\ We also wish to point out that by including the possibility of condensation of certain operators, we are looking at the GZ dynamics w.r.t.~a dynamically improved vacuum, in particular an improved calculation of the effective action, and thus of the horizon condition via \eqref{gapgamma}, becomes possible.

\subsection{The action with inclusion of $d=2$ condensates}
We propose to study the following extended action,
\begin{eqnarray}\label{CGZ}
\Sigma_\CGZ &=& S_\GZ  + S_{A^2} + S_{\varphi \overline \varphi}  + S_{\overline \omega \omega} + S_{\overline{\varphi} \overline \varphi, \overline \omega \overline \varphi }  + S_{\varphi \varphi, \omega \varphi } + S_\vac
\end{eqnarray}
whereby $S_\GZ$ is given by equation \eqref{GZaction} and
\begin{eqnarray}
S_{A^2} &=& \int \d^d x \left( \frac{\tau }{2}A_{\mu }^{a}A_{\mu }^{a} - \frac{\zeta }{2}\tau ^{2}\right)\;, \nonumber\\
S_{\varphi \overline \varphi } & =&  \int \d^4 x\; s(  P \overline \varphi^a_i \varphi^a_i ) ~=~  \int \d^4 x \left[  Q  \overline \varphi^a_i \varphi^a_i- P  \overline \varphi^a_i \omega^a_i\right]  \;, \nonumber\\
S_{\overline \omega \omega } & =&  \int \d^4 x\; s(  V \overline \omega^a_i \omega^a_i) ~=~  \int \d^4 x \left[ W\overline \omega^a_i \omega^a_i - V \overline \varphi^a_i \omega^a_i \right]  \;, \nonumber\\
S_{\overline{\varphi} \overline \varphi, \overline \omega \overline \varphi } & =&  \frac{1}{2} \int \d^4 x\; s(  \overline G^{ij} \overline \omega^a_i \overline \varphi^a_j) ~=~  \int \d^4 x \left[  \overline H^{ij} \overline \omega^a_i \overline \varphi^a_j + \frac{1}{2} \overline G^{ij} \overline \varphi^a_i \overline \varphi^a_j \right]  \;, \nonumber\\
S_{\varphi \varphi, \omega \varphi } &=& \frac{1}{2} \int \d^4 x\; s(  H^{ij} \varphi^a_i  \varphi^a_j) ~=~   \int \d^4 x  \left[ \frac{1}{2} G^{ij} \varphi^a_i \varphi^a_j - H^{ij} \omega^a_i \varphi^a_j \right] \;, \nonumber\\
S_\vac &=&  \int \d^4 x \left[ \kappa (G^{ij} \overline G^{ij} - 2 H^{ij} \overline H^{ij}) +  \lambda (G^{ii} \overline G^{jj} - 2 H^{ii} \overline H^{jj}) \right] \nonumber\\
 && - \int \d^4 x \left[ \alpha   (Q Q +  Q W) + \beta ( QW + W W)  + \chi Q \tau + \delta W \tau \right]\;.
\end{eqnarray}
We have introduced a source $\tau$ and 4 new doublets of sources, i.e.
\begin{eqnarray}
s \tau &=& 0 \;, \nonumber\\
s P &=& Q \;,   \qquad sQ=0 \;, \nonumber\\
s V &=& W \;,  \qquad sW =0 \; \nonumber\\
s  \overline G^{ij} &=& 2 \overline H^{ij} \;,  \qquad s\overline H^{ij} =0 \; \nonumber\\
s H^{ij} &=& G^{ij}\;,  \qquad s G^{ij} = 0 \;
\end{eqnarray}
whereby $\tau$ is a bosonic source and $P$, $V$, $H^{ij}$ and $\overline H^{ij}$   are Grassmann quantities. For consistency, the sources with double index $^{ij}$   are symmetric in these indices. In this light, we use the following definition for the derivative w.r.t.~a symmetric source $\Lambda_{kl}$:
\begin{equation}\label{der}
    \frac{\delta \Lambda_{ij}}{\delta
    \Lambda_{k\ell}}=\frac{1}{2}\left(\delta_{ik}\delta_{j\ell}+\delta_{i\ell}\delta_{jk}\right)\;.
\end{equation}
Notice that some sources have double indices, e.g.~$H^{ij}$, while other sources have no indices, e.g.~$P$. The reason for this is only related to the algebraic proof of the renormalization in order to keep certain symmetries, and has no further meaning. \\
\\
We have also introduced a vacuum term, $S_\vac$, which shall be important for the renormalization of the vacuum energy. As shown in \cite{Verschelde:2001ia,Verschelde:1995jj}, the dimensionless LCO parameters $\alpha $, $\beta$, $\chi$, $\delta$ and $\zeta$ of the quadratic terms in the sources are needed to account for the divergences present in the correlation functions like
$\braket{\mathcal O_i(k) \mathcal O_j(-k) }$, with $\mathcal O_i$ one of the operators given in expression \eqref{operators}.\\
\\
Now we can prove that the action \eqref{CGZ} is renormalizable to all orders. The proof is very similar to  that of the renormalizability of the GZ action, the only difficulty is that  the  mixing between different sources and parameters is now allowed. We refer to the appendices \ref{appendixB} and \ref{appendixC} for all the details. \\
\\
For the rest of the  work, we are only interested in a  restricted number of condensates.  Therefore, we first set the source $W =0$, which is coupled to $\overline \omega \omega$, as this is not of our current interest\footnote{There is no quadratic coupling of $\omega$ and $\overline \omega$ to the gluon sector, thus such a condensate would not directly influence the gluon propagator.}, and we also set $P = V = \eta = 0$, as we have introduced these  sources only to preserve the BRST symmetry. Secondly, we also take $H^{ij} = \overline H^{ij} = 0$ and we set $G^{ij} = \delta^{ij} G$ and $\overline G^{ij} = \delta^{ij} \overline G$. The action \eqref{CGZ} becomes,
\begin{eqnarray}\label{startxx}
\Sigma_\CGZ &=& S_\GZ + \int \d^4 x  \left[ Q  \overline \varphi^a_i \varphi^a_i   + \frac{1}{2} \tau A_{\mu }^{a}A_{\mu }^{a}  - \frac{1}{2}\zeta \tau ^{2}  - \alpha   Q Q    - \chi Q \tau\right] \nonumber\\
&& + \int \d^4 x \left[ \frac{1}{2} \overline G \overline \varphi^a_i \overline \varphi^a_i + \frac{1}{2} G \varphi^a_i \varphi^a_i  + \varrho G \overline G \right]\;,
\end{eqnarray}
whereby $(\kappa d (N^2 -1) + \lambda d^2(N^2 -1)^2)$ was replaced by one parameter $\varrho$.

\subsection{A diagrammatical look at  the potential mixing and  at the vacuum divergences\label{diagrammatical}}
\begin{figure}[H]
   \centering
       \includegraphics[width=5cm]{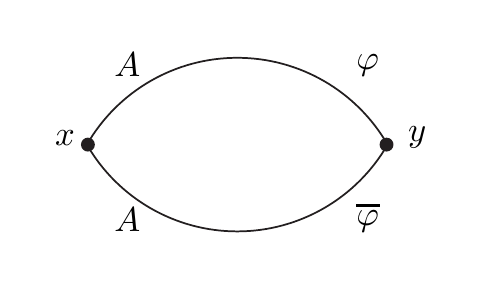}
   \caption{1-loop diagram for $\braket{A^2 (x) \overline \varphi \varphi (y)}$.}
   \label{1loopd}
\end{figure}

Before starting the calculation of the effective action, we can provide some simplification  with the help of a diagrammatical argument. Firstly, looking at the action \eqref{startxx}, we see that a term $\chi Q \tau$ is present. This term is responsible for killing the divergences in the vacuum correlators $\Braket{A^2 (x) \overline \varphi \varphi (y)}$ for $x\to y$. However, we can prove that there are no divergences of this kind in  the one loop diagrams. Let us  start by considering  these one loop diagrams. There is only one possible type of diagram for $\Braket{A^2 (x) \overline \varphi \varphi (y)}$, as displayed in Figure \ref{1loopd}.

\noindent The UV behavior of this diagram is finite, as can be extracted from the list of propagators \eqref{summarypropGZ}. Indeed, for large momenta, the corresponding integral of the diagram \eqref{1loopd} behaves like $\sim \int \d^4 p \frac{1}{p^4}\frac{1}{p^4}$, which is perfectly finite  in the UV. Therefore, $ \lim_{x \to y}\Braket{A^2 (x) \overline \varphi \varphi (y)}$ is not divergent at one loop. In the next section, we shall explicitly prove this.\\
\\
At two loops, it is not possible to present the same argument as there exists a diagram which can be logarithmically divergent:

\begin{figure}[H]
   \centering
       \includegraphics[width=4.5cm]{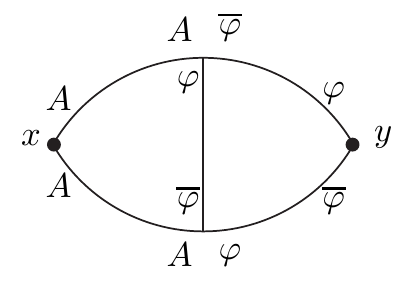}
   \caption{A possible divergent 2-loop diagram for $\braket{A^2 (x) \overline \varphi \varphi (y)}$.}
   \label{1loopd2}
\end{figure}

\noindent as can be checked from the list of propagators \eqref{summarypropGZ}.\\
\\
Secondly, we can also have a look at the mixing of the operators $A^2$ and $\overline \varphi \varphi$. In the algebraic analysis, see appendix \ref{appendixC}, we have found that a mixing is possible between the different operators, see equation \eqref{mixing}. This means that algebraically, a counterterm  of the type $Q A_\mu A_\mu$ is allowed. This counterterm is needed to cancel the infinities of the following type of diagrams:

\begin{figure}[H]
   \centering
       \includegraphics[width=4.5cm]{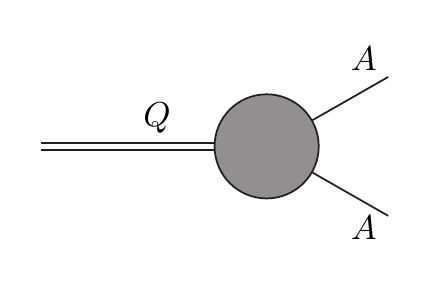}
   \label{mixxx}
\end{figure}

\noindent However, we can prove that there are no infinities at one loop, as the only possible diagram is given by,

\begin{figure}[H]
   \centering
       \includegraphics[width=5.5cm]{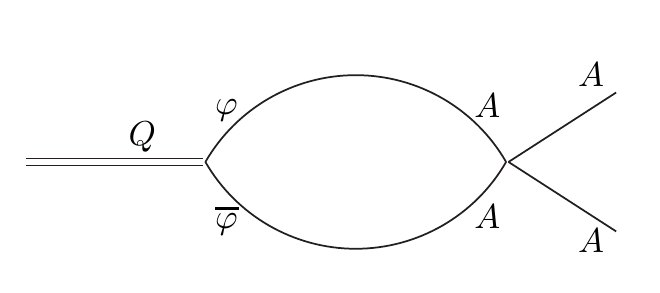}
   \label{mixx2}
\end{figure}

\noindent which is similar to the diagram in Figure \ref{1loopd}. We can thus conclude that the mixing can only start at two loops. Again, we cannot exclude divergences at two loops, due to a similar diagram as in Figure \ref{1loopd2}.

\section{The effective action}
In this section, we shall try to calculate the effective action. The calculation is quite technical and  shall therefore be split in different steps, although the result is reasonably compact and can be immediately found in expression \eqref{finaleffective}.\\
\\
The energy functional can be written as
\begin{eqnarray}\label{enfunc}
\e^{- W(Q, \tau, G, \overline G)} &=& \int [\d A_\mu][\d c][\d\overline c][\d b] [\d \varphi] [\d\overline \varphi] [\d \omega] [\d \overline \omega] \e^{-\Sigma_\CGZ}\;,
\end{eqnarray}
with $\Sigma_\CGZ$ given by equation \eqref{startxx}. We recall that in $d = 4 -\epsilon$ dimensions, we have the following dimensionalities,
\begin{eqnarray}
\left[A_\mu\right] &=& \left[\varphi \right] =  \frac{d-2}{2} = 1- \frac{\epsilon}{2}\;, \nonumber\\
\left[g\right] &=& \frac{4-d}{2} = \frac{\epsilon}{2}\;, \nonumber\\
\left[\tau \right] &=& \left[Q\right] = \left[G\right] = \left[\overline G\right] = 2   \;, \nonumber\\
\left[\zeta\right] &=& \left[\alpha\right]  =  \left[\chi\right] = \left[\varrho\right] = d-4 = -\epsilon\;.
\end{eqnarray}

\subsection{The LCO formalism}
In order to calculate the effective action, we shall follow the local composite operator (LCO) formalism developed  in \cite{Verschelde:2001ia,Verschelde:1995jj}. Let us outline the main idea. We start from a LCO $\mathcal O$, in our case a local dimension two operator within a dimension four theory. As  done several times, we couple the operator(s) of interest to an appropriate source(s) $J$, and add  the term $J \mathcal O$ to the Lagrangian. This gives rise to a functional $W (J)$  which we need to Legendre transform to find the effective potential. However, as already  observed, novel infinities shall arise, which are proportional to $J^2$.  These  infinities are due to the divergences in the correlator $ \lim_{ x \to y}\braket{\mathcal O(x) \mathcal O(y)}$, as  explained in section \ref{diagrammatical}. Therefore, in general, a term proportional to $J^2$ is always needed in the counterterm, and the starting action needs  to display a  term\footnote{For an example, see the action \eqref{startxx}, where the term $ - \frac{1}{2}\zeta \tau ^{2}  - \alpha   Q Q    - \chi Q \tau$ is needed in the starting action. The sources $Q$ and $\tau$ are coupled to the LCO operators $\mathcal O_1 = \overline \varphi_i \varphi_i $ and $\mathcal O_2 = A_\mu A_\mu$. Note that here, also a mixing term $\chi Q \tau$ accounting for the divergences in $\lim_{x \to y}\braket{\mathcal O_1(x) \mathcal O_2 (y)}$ is present.  } $\zeta J^2$.   The novel parameter $\zeta$,   called the LCO parameter,  is needed to absorb the divergences in $J^2$, i.e.~$\delta \zeta J^2$.  With the inclusion of the term $\zeta J^2$, the functional $W(J)$ obeys the following homogeneous RGE
\begin{equation}\label{RGEplus}
\left( \mu\frac{ \p}{\p \mu} + \beta(g^2) \frac{\p}{\p g^2} - \gamma_J(g^2) \int \d^4 x J \frac{\delta}{\delta J} + \eta(g^2 , \zeta) \frac{\p}{\p \zeta}  \right)W(J) = 0\;,
\end{equation}
with $\eta(g^2, \zeta)$ the running of $\zeta$,
\begin{equation}
\mu \frac{\p}{\p \mu} \zeta = \eta (g^2 , \zeta)\;.
\end{equation}
Notice that it is necessary to include the running of $\zeta$ at this point.\\
\\
Now the question is, how can we determine this seemingly arbitrary parameter $\zeta$? This is possible by employing the renormalization group equations. We can write
\begin{equation}
\zeta_0 J_0^2 = \mu^{-\epsilon} (\zeta J^2 + \delta \zeta J^2)\;,
\end{equation}
whereby the second term of the r.h.s.~represents the counterterm. As the l.h.s.~is independent from $\mu$, we can derive both sides w.r.t.~$\mu$ to find:
\begin{equation}\label{diffeq}
- \epsilon  (\zeta + \delta \zeta) + \left( \mu \frac{\p}{\p \mu} \zeta + \mu \frac{\p}{\p \mu} (\delta \zeta) \right) - 2 \gamma_J(g^2) (\zeta + \delta \zeta) = 0\;,
\end{equation}
whereby $\gamma_J(g^2)$ is the anomalous dimension of $J$. As we can consider $\zeta$ to be a function of $g^2$, and by evoking the $\beta$ function,
\begin{equation}
\beta (g^2) = \mu \frac{\p}{\p \mu} g^2
\end{equation}
the equation \eqref{diffeq} becomes,
\begin{equation}\label{diffeqx}
\beta(g^2)  \frac{\p}{\p g^2} \zeta (g^2) =   2 \gamma_J(g^2) \zeta +  f(g^2)\;.
\end{equation}
with $f(g^2) =  \epsilon  \delta \zeta  - \beta(g^2)  \frac{\p}{\p g^2} (\delta \zeta)  + 2 \gamma_G(g^2)  \delta \zeta $. The general solution of this differential equation reads
\begin{equation}
\zeta(g^2) = \zeta_p(g^2) + \alpha \exp \left( 2\int_1^{g^2}  \frac{\gamma_J(z)}{\beta (z)} \d z\right)\;,
\end{equation}
with $\zeta_p(g^2)$ a particular solution of \eqref{diffeqx}. A possible particular solution is given by
\begin{equation}
\zeta_p(g^2) = \frac{c_0}{g^2} + c_1 \hbar + c_2 g^2 \hbar^2 + \ldots \;.
\end{equation}
whereby we have temporarily introduced the dependence on $\hbar$. Notice therefore that the $n$-loop result for $\zeta(p^2)$ will require the $(n+ 1)$ loop results of $\beta(g^2)$, $\gamma_J(g^2)$ and $f(g^2)$. As we would like $\zeta$ to be multiplicatively renormalizable, we set $\alpha = 0$. In this case we have that
\begin{equation}
\zeta(g^2) + \delta \zeta(g^2) = \zeta_0 = Z_\zeta \zeta (g^2)\;,
\end{equation}
and we  have removed  the independent parameter $\alpha.$ Also, now that $\zeta$ is a function of $g^2$, the RGE \eqref{RGEplus} becomes
\begin{equation}
\left( \mu\frac{ \p}{\p \mu} + \beta(g^2) \frac{\p}{\p g^2} - \gamma_J(g^2) \int \d^4 x J \frac{\delta}{\delta J}  \right)W(J) = 0\;,
\end{equation}
as deriving w.r.t.~$\zeta$ is now incorporated in deriving w.r.t.~$g^2$.\\
\\
After determining the LCO parameter $\zeta$, the next step is to calculate the effective action by doing a Legendre transformation. However, it shall be easier to perform a Hubbard-Stratonovich transformation on $W(J)$, whereby we introduce an auxiliary field $\sigma$ describing the composite operator $\mathcal O$. In this way, we  can get rid of the quadratic term in $J^2$ and a clear relation with the effective action emerges,  as it will be shown later on in this section. We only need to mention that the case we are handling here is a bit more complicated due to the mixing of the operators $\mathcal O_1 = \overline \varphi_i \varphi_i $ and $\mathcal O_2 = A_\mu A_\mu$, and  to the mixing of the vacuum divergences. However, the basic principles  remain  the same.

\subsection{Differential equation for the LCO parameters $\zeta$, $ \alpha$, $ \chi$ and $\varrho$}
We shall try to determine the four LCO parameters $\zeta$, $\alpha$, $\chi$ and $\varrho$. We shall first derive a differential equation for these parameters, in an analogous way as in \cite{Verschelde:2001ia,Dudal:2009tq}. As there can be mixing, we shall define $\delta \zeta$, $\delta \omega$ and $\delta \chi$ as follows
\begin{equation}\label{notatie}
-\frac{1}{2} \zeta_0 \tau_0^2 - \alpha_0 Q_0^2 - \chi_0 Q_0 \tau_0  = -\mu ^{-\epsilon} \left( \frac{1}{2} \zeta \tau^2 + \alpha Q^2 + \chi Q \tau + \frac{1}{2} \delta \zeta \tau^2 + \delta \alpha Q^2 +\delta \chi Q \tau \right)\;,
\end{equation}
while $\delta \varrho$ can be defined independently:
\begin{equation}\label{nata2}
\varrho_0 G_0 \overline G_0 =  \mu ^{-\epsilon}  Z_\varrho Z_G Z_{\overline G} \varrho G \overline G =  \mu ^{-\epsilon}  \left(1 + \frac{\delta \varrho}{\varrho} \right) \varrho G \overline G \;.
\end{equation}
We further define the anomalous dimension of $G$,
\begin{equation}\label{anoG}
\mu \frac{\p}{\p \mu} \ln Z_G   = \gamma_G(g^2) \quad \Rightarrow \quad \mu \frac{\p}{\p \mu} G = - \gamma_G(g^2)G \;,
\end{equation}
which is exactly the same as the anomalous dimension of $\overline G$ as $Z_G = Z_{\overline G}$. To define the anomalous dimensions of $Q$ and $\tau$, we start from equation \eqref{mixing}:
\begin{eqnarray}
  \underbrace{\begin{bmatrix}
     Q_0 \\
     \tau_0
  \end{bmatrix}}_{X_0}
 &= &
          \underbrace{\begin{bmatrix}
            Z_{QQ} &0 \\
           Z_{\tau Q} &  Z_{\tau\tau}          \end{bmatrix}}_{Z}
\underbrace{\begin{bmatrix}
    Q\\
    \tau
  \end{bmatrix}}_{X}\,,
\end{eqnarray}
a relation stemming from the algebraic renormalization.  To the matrix $Z$, we can associate the anomalous dimension matrix $\Gamma$:
\begin{equation}
\mu \frac{\p}{\p \mu} Z = Z \Gamma\;,
\end{equation}
and thus
\begin{equation}\label{gammas}
\Gamma = Z^{-1} \mu \frac{\p}{\p \mu} Z = \begin{bmatrix}
              Z_{QQ}^{-1} \mu \frac{\p}{\p \mu} Z_{QQ} & 0 \\
              -Z_{\tau Q} \mu \frac{\p}{\p \mu} Z_{QQ} + Z_{\tau \tau}^{-1} \mu \frac{\p}{\p \mu} Z_{\tau Q} &  Z_{\tau\tau}^{-1} \mu \frac{\p}{\p \mu} Z_{\tau \tau}          \end{bmatrix} =  \begin{bmatrix}
              \gamma_{QQ} & 0 \\
              \Gamma_{21} &  \gamma_{\tau \tau}          \end{bmatrix}\;.
\end{equation}
This matrix is then related to the anomalous dimension of the operators:
\begin{align}
X_0 &= Z X &\Rightarrow 0 &= \mu \frac{\p Z}{ \p \mu}  X + Z \mu \frac{\p X}{\p \mu} &\Rightarrow \mu \frac{\p X}{\p \mu} &= - \Gamma X \,,
\end{align}
so the anomalous dimensions of the sources $Q$ and $\tau$ is given by
\begin{equation}
\mu \frac{\p}{\p \mu} \begin{bmatrix} Q \\ \tau \end{bmatrix} = \begin{bmatrix} -\gamma_{QQ} & 0  \\ - \Gamma_{21} & - \gamma_{\tau\tau} \end{bmatrix} \begin{bmatrix} Q \\ \tau \end{bmatrix}\;.
\end{equation}
With these definitions in mind, we can derive a differential equation for $\delta \zeta$, $\delta \omega$, $\delta \chi$ and $\delta \varrho$. We start with that of $\delta \varrho$. Starting from expression \eqref{nata2} and deriving w.r.t.~$\mu$, we find
\begin{equation}
- \epsilon  (\varrho + \delta \varrho) + \left( \mu \frac{\p}{\p \mu} \varrho + \mu \frac{\p}{\p \mu} (\delta \varrho) \right) - 2 \gamma_G(g^2) (\varrho + \delta \varrho) = 0\;.
\end{equation}
As we can consider $\varrho$ to be a function of $g^2$, according to the standard LCO formalism, we can rewrite this equation as
\begin{equation}
\beta(g^2)  \frac{\p}{\p g^2} \varrho (g^2) =  \epsilon (\varrho + \delta \varrho) - \beta(g^2)  \frac{\p}{\p g^2} (\delta \varrho)  + 2 \gamma_G(g^2) (\varrho + \delta \varrho) \;.
\end{equation}
As $\varrho$ is finite, we can even further simplify this into
\begin{equation}\label{diff2}
\beta(g^2)  \frac{\p}{\p g^2} \varrho (g^2) =   2 \gamma_G(g^2) \varrho + \epsilon \delta \varrho - \beta(g^2)  \frac{\p}{\p g^2} (\delta \varrho)  + 2 \gamma_G(g^2)  \delta \varrho \;.
\end{equation}
In an analogous fashion, we can find  the differential equations for $\delta \zeta$, $\delta \omega$ and $\delta \chi$. If we derive \eqref{notatie} w.r.t.~$\mu$, we find the following set of coupled differential equations
\begin{align}\label{diff1}
&  \beta(g^2)  \frac{\p}{\p g^2} \frac{\zeta (g^2)}{2}  =  \frac{\epsilon}{2}  \delta \zeta  - \frac{1}{2}\beta(g^2)  \frac{\p}{\p g^2} (\delta \zeta)  +  \gamma_{\tau \tau}(g^2) (\zeta + \delta \zeta) \;,\nonumber\\
& \beta(g^2)  \frac{\p}{\p g^2} \alpha (g^2)  =  \epsilon \delta \alpha  - \beta(g^2)  \frac{\p}{\p g^2} (\delta \alpha)  + 2  \gamma_{QQ}(g^2) (\alpha + \delta \alpha) + \Gamma_{21}(g^2) (\chi + \delta \chi) \;, \nonumber\\
& \beta(g^2)  \frac{\p}{\p g^2} \chi (g^2) =  \epsilon  \delta \chi  - \beta(g^2)  \frac{\p}{\p g^2} (\delta \chi)  +   \gamma_{QQ}(g^2) (\chi+ \delta \chi) + \gamma_{\tau \tau}(g^2) (\chi + \delta \chi)  + \Gamma_{21} (g^2) (\zeta + \delta \zeta) \;.
\end{align}

\subsection{Determination of the LCO parameters $\delta \zeta$, $\delta \alpha$, $\delta \chi$ and $\delta \varrho$ \label{section4.2}}
In order to determine the counterterm parameters $\delta \zeta$, $\delta \alpha$, $\delta \chi$ and $\delta \varrho$ at one loop, we need to calculate the one loop divergence of the energy functional $W(Q, \tau, G, \overline G)$. The details of these calculations can be found in appendix \ref{detailsvery}. From section \ref{diagrammatical}, we know that at one loop, $\delta \chi$ should be zero. This observation shall serve as a check  of our computations.\\
\\
In the appendix \ref{detailsvery}, equation \eqref{deltas}, we have found
\begin{eqnarray}\label{deltasx}
\delta \zeta &=& -\frac{1}{\epsilon} \frac{3}{16 \pi^2} (N^2-1) \;,\nonumber\\
\delta \alpha &=& -\frac{1}{\epsilon} \frac{1}{4 \pi^2} (N^2-1)^2 \;, \nonumber\\
\delta \chi &=& 0 \;,\nonumber\\
\delta \varrho &=& \frac{1}{\epsilon} \frac{1}{4 \pi^2} (N^2-1)^2 \;.
\end{eqnarray}
 The value of $\delta \zeta$ provides already a first check of our results. In fact, this quantity has been calculated up to three loops, see \cite{Verschelde:2001ia,Browne:2003uv}.   Our  one loop value for $\delta \zeta$ coincides with that reported in \cite{Verschelde:2001ia,Browne:2003uv}. Secondly, we also see that indeed $\delta \chi = 0$ at one loop, which nicely confirms our diagrammatical power counting argument.

\subsection{Solving the differential equations for $\zeta$, $\alpha$, $\chi$ and $\varrho$}
In this section, we shall try to solve the differential equations \eqref{diff2} and \eqref{diff1}, when possible. For these calculations, it is useful to keep in mind the $\beta$ function, here given up to two loops
\begin{eqnarray}
\beta(g^2) &=& -\epsilon g^2 - 2\left(  \beta_0 g^4 + \beta_1 g^6 + O(g^8)  \right)\;,
\end{eqnarray}
with
\begin{eqnarray}
\beta_0 &=&  \frac{11}{3}  \left( \frac{N}{16 \pi^2} \right)\;,\qquad \beta_1 ~=~  \frac{34}{3}                \left( \frac{N}{16 \pi^2}\right)^2\;,
\end{eqnarray}
in order to keep track of the orders.\\
\\
We start with \eqref{diff2},
\begin{equation}\label{diffa}
\beta(g^2)  \frac{\p}{\p g^2} \varrho (g^2) =   2 \gamma_G(g^2) \varrho + \epsilon \delta \varrho - \beta(g^2)  \frac{\p}{\p g^2} (\delta \varrho)  + 2 \gamma_G(g^2)  \delta \varrho \;.
\end{equation}
In order to solve this differential equation, we need to parameterize $\varrho$ as follows:
\begin{equation}
\varrho = \frac{\varrho_0}{g^2} + \varrho_1 + \varrho_2 g^2 + \mathcal O(g^4)\;.
\end{equation}
We also need the explicit value of the anomalous dimension $\gamma_G$. We have from the definition \eqref{anoG} that
\begin{equation}
 \gamma_G(g^2)= \mu \frac{\p}{\p \mu} \ln Z_G \;,
\end{equation}
and thus we need the value of $Z_G$. From the renormalization factors \eqref{Zfac} and \eqref{Z3}, we find that
\begin{equation}
 \gamma_G(g^2)= - \mu \frac{\p}{\p \mu} \ln Z_\varphi =  - \mu \frac{\p}{\p \mu} \ln (Z_g^{-1} Z_A^{-1/2}) \;.
\end{equation}
In \cite{Gracey:2002yt}, the factors $Z_g$ and $Z_A$ have been calculated up to three loops,
\begin{eqnarray}
Z_A &=& 1+ \frac{13}{6} \frac{1}{\epsilon} \frac{N g^2}{16\pi^2} + \left(\frac{-13}{8} \frac{1}{\epsilon^2} + \frac{59}{16} \frac{1}{\epsilon} \right) \left( \frac{N g^2}{16\pi^2} \right)^2  + \ldots \;,\nonumber\\
Z_g &=& 1- \frac{11}{6} \frac{1}{\epsilon} \frac{N g^2}{16\pi^2} + \left(\frac{121}{24} \frac{1}{\epsilon^2} - \frac{17}{6} \frac{1}{\epsilon} \right) \left( \frac{N g^2}{16\pi^2} \right)^2 +\ldots \;.
\end{eqnarray}
So one can calculate $\gamma_G(g^2)$ up to three loops if necessary. Here only the first loop shall be useful for our calculations, i.e.
\begin{equation}
\gamma_G(g^2) = \frac{3}{4} \frac{N g^2}{16 \pi^2} + \ldots \;,
\end{equation}
as $\delta \varrho$, see equation \eqref{deltas}, is only known up to lowest order. With this information, we can solve the differential equation \eqref{diffa} up to lowest order, by matching the corresponding orders in $g^2$
\begin{equation}
\varrho = \frac{24}{53} \frac{ (N^2 - 1)^2}{N g^2} + \varrho_1 + \ldots \;.
\end{equation}
Unfortunately, we cannot solve the differential equation for $\varrho_1$ as we would require the two loop value of $\delta \varrho$, which is however not easily computed. Therefore, in the current work, we leave this value as a  parameter to be determined. \\
\\
Let us now turn to the set of differential equations \eqref{diff1}. We can do a similar analysis as above for the first differential equation, namely
\begin{equation}
\beta(g^2)  \frac{\p}{\p g^2} \frac{\zeta (g^2)}{2}  =  \frac{\epsilon}{2}  \delta \zeta  - \frac{1}{2}\beta(g^2)  \frac{\p}{\p g^2} (\delta \zeta)  +  \gamma_{\tau \tau}(g^2) (\zeta + \delta \zeta)\;.
\end{equation}
We shall again parameterize $\zeta$ as follows:
\begin{equation}\label{opl1}
\zeta = \frac{\zeta_0}{g^2} + \zeta_1 + \zeta_2 g^2 + \mathcal O(g^4)\;.
\end{equation}
In fact, we can even solve this differential equation to two loops. From \cite{Verschelde:2001ia,Browne:2003uv,Dudal:2009tq}, we know that
\begin{multline}
\delta \zeta = \frac{N^2 - 1}{16 \pi^2} \Biggl[ -\frac{3}{\epsilon} + \left( \frac{35}{2} \frac{1}{\epsilon^2} -  \frac{139}{6} \frac{1}{\epsilon} \right)\left(  \frac{g^2 N}{16\pi^2} \right)   \\ +\left(-\frac{665}{6} \frac{1}{\epsilon^3} + \frac{6629}{36} \frac{1}{\epsilon^2} - \left(  \frac{71551}{432} + \frac{231}{16} \zeta(3) \right)  \frac{1}{\epsilon} \right)\left(  \frac{g^2 N}{16\pi^2} \right)^2 \Biggr]\;,
\end{multline}
and
\begin{multline}
 Z_{\tau\tau} = 1- \frac{35}{6} \frac{1}{\epsilon} \left( \frac{g^2 N }{16 \pi^2}\right)  + \left[ \frac{2765}{72}  \frac{1}{\epsilon^2}-\frac{449}{48}  \frac{1}{\epsilon}   \right] \left( \frac{g^2 N }{16 \pi^2}\right)^2 \\+ \left[ -\frac{113365}{432}\frac{1}{\epsilon^3} +  \frac{41579}{576}  \frac{1}{\epsilon^2}+\left(-\frac{75607}{2592}-\frac{3}{16}\zeta(3)\right)\frac{1}{\epsilon}  \right] \left( \frac{g^2 N }{16 \pi^2}\right)^3\;,
\end{multline}
so that from \eqref{gammas}
\begin{eqnarray}\label{ress1}
\gamma_{\tau\tau}(g^2) =  \frac{35}{6}\left(  \frac{g^2 N}{16\pi^2} \right) +  \frac{449}{24}\left(  \frac{g^2 N}{16\pi^2} \right)^2 +  \left( \frac{94363}{864} + \frac{9}{16} \zeta(3) \right)\left(  \frac{g^2 N}{16\pi^2} \right)^3\;.
\end{eqnarray}
By solving the differential equation for $\zeta$, we can determine $\zeta$ to one loop order. In principle, we can even go one loop further with the known results. However, as we shall only determine the effective potential to one loop order, we do not need this next loop result. We find,
\begin{eqnarray}\label{weljui}
\zeta &=& \frac{N^2 -1}{16 \pi^2} \left[\frac{9}{13} \frac{16 \pi^2}{ g^2 N} + \frac{161}{52} \right]\;,
\end{eqnarray}
see also \cite{Dudal:2009tq}. \\
\\
The second and third differential equation of \eqref{diff1} are coupled.  However, they can be simplified and decoupled as $\delta \chi = 0$:
\begin{eqnarray}
\beta(g^2)  \frac{\p}{\p g^2} \alpha (g^2)  &=&  2  \gamma_{QQ}(g^2) \alpha + \epsilon  \delta \alpha  - \beta(g^2)  \frac{\p}{\p g^2} (\delta \alpha)  + 2  \gamma_{QQ}(g^2) \delta \alpha + \Gamma_{21}(g^2) \chi  \;,\nonumber\\
\beta(g^2)  \frac{\p}{\p g^2} \chi (g^2) &=&      \gamma_{QQ}(g^2) \chi + \gamma_{\tau \tau}(g^2) \chi  + \Gamma_{21} (g^2) (\zeta + \delta \zeta) \;.
\end{eqnarray}
Fortunately, we know that $\Gamma_{21} = 0$ at lowest order, from the diagrammatical argument in section \ref{diagrammatical}. Therefore, we can set $\Gamma_{21} = 0 + O(g^4)$. When parameterizing as usual
\begin{align}
\alpha &= \frac{\alpha_0}{g^2} + \alpha_1 + \alpha_2 g^2 + \mathcal O(g^4) \;, &  \chi &= \frac{\chi_0}{g^2} + \chi_1 + \chi_2 g^2 + \mathcal O(g^4) \;,
\end{align}
we find for the solution of the differential equations
\begin{eqnarray}\label{opl2}
\alpha_0 &=& -\frac{24  (N^2-1)^2}{35 N} \;,\nonumber\\
\chi_0 &=& 0 \;.
\end{eqnarray}

\subsection{Hubbard-Stratonovich transformations}
In this section, we shall get rid of the unwanted quadratic source dependence by the introduction of multiple Hubbard-Stratonovich (HS) fields.
We can then rewrite the relevant part of the action in terms of finite fields and sources:
\begin{eqnarray*}
 && \int \d^4 x  \Bigl[ \underbrace{ Z_{QQ} Z_\varphi }_{c} Q \overline \varphi^a_i \varphi^a_i   + \underbrace{ \frac{1}{2} Z_A Z_{\tau\tau}}_{b} \tau A_{\mu }^{a}A_{\mu }^{a}   + \underbrace{ \frac{1}{2} Z_A Z_{\tau Q }}_{a} Q A_{\mu }^{a}A_{\mu }^{a}  - \underbrace{ \frac{1}{2} Z_{\zeta \zeta} Z_{\tau\tau}^2 \zeta }_{\zeta'} \mu^{-\epsilon} \tau ^{2} \nonumber\\
 && -\underbrace{ Z_{QQ}^2 Z_{\alpha\alpha} \alpha }_{\alpha'} \mu^{-\epsilon} Q Q    - \underbrace{ Z_{QQ} Z_{\chi \chi} Z_{\tau\tau}  \chi}_{\chi'}  \mu^{-\epsilon} Q \tau\Bigr]\nonumber\\
 &&+ \int \d^4 x \Bigl[  Z_{ G} Z_\varphi \frac{1}{2} \overline G \overline \varphi^a_i \overline \varphi^a_i + Z_G Z_\varphi \frac{1}{2} G \varphi^a_i \varphi^a_i  + Z_\varrho Z_G^2 \varrho G \overline G\Bigr]\;.
\end{eqnarray*}
We shall now perform the following Hubbard-Stratonovich (HS) transformations by multiplying expression \eqref{enfunc} with the following unities\footnote{We dropped irrelevant normalization factors.},
\begin{eqnarray}\label{unities}
1 &=& \int [\d \sigma_1] \e^{-\frac{1}{4 \zeta'} \int\d^d x \left( \frac{\sigma_1 }{g}  + b \mu^{\epsilon/2} A^2 - 2 \zeta' \mu^{-\epsilon/2} \tau  - \chi' \mu^{-\epsilon/2} Q  \right)^2 }\;, \nonumber\\
1 &=& \int [\d \sigma_2] \e^{- \frac{1}{4 \zeta' [4 \alpha' \zeta' - \chi^{\prime 2}  ] } \int\d^d x \left( \frac{\sigma_2 }{g}  +  (b \chi' - 2 a \zeta') \mu^{\epsilon/2} A^2 - 2 c \zeta' \mu^{\epsilon/2} \overline \varphi \varphi + (4 \alpha' \zeta' - \chi^{\prime 2}) \mu^{-\epsilon/2} Q    \right)^2 } \;, \nonumber\\
1 &=& \int [\d \sigma_3] \e^{-\frac{1}{ 4 Z_\varrho Z_G^2  \varrho } \int\d^d x \left( \frac{\sigma_3 }{g}  + \frac{1}{2} \mu^{\epsilon/2} Z_{ G} Z_\varphi \overline \varphi \overline \varphi  + \frac{1}{2} \mu^{\epsilon/2} Z_{ G} Z_\varphi  \varphi \varphi + Z_{ G}^2 Z_\varrho \varrho \mu^{-\epsilon/2} \overline G + Z_G^2 Z_\varrho \varrho \mu^{-\epsilon/2} G   \right)^2 } \;, \nonumber\\
1 &=& \int [\d \sigma_4] \e^{-\frac{1}{ 4 Z_\varrho Z_G^2  \varrho } \int\d^d x \left( \frac{ \sigma_4 }{g}  + \frac{\ii}{2} \mu^{\epsilon/2} Z_{ G} Z_\varphi \overline \varphi \overline \varphi  -\frac{\ii}{2} \mu^{\epsilon/2} Z_{ G} Z_\varphi  \varphi \varphi - \ii Z_{ G}^2 Z_\varrho \varrho \mu^{-\epsilon/2} \overline G + \ii Z_G^2  Z_\varrho \varrho \mu^{-\epsilon/2} G   \right)^2 }  \;,
\end{eqnarray}
whereby we have introduced four new fields, $\sigma_1$,$\sigma_2$, $\sigma_3$ and $\sigma_4$. By doing these HS transformations, we can remove the quadratic sources and rewrite the functional energy as
\begin{multline}\label{enfunc2}
\e^{- W(Q, \tau, G, \overline G)} = \int [\d A_\mu][\d c][\d\overline c][\d b] [\d \sigma_1][\d \sigma_2][\d \sigma_3] [\d \sigma_4] [\d \varphi] [\d\overline \varphi] [\d \omega] [\d \overline \omega]\\
\times \e^{\left[ - \int\d^d x \left( \mathcal L(\phi, \sigma_1, \ldots,\sigma_4)   - \mu^{-\epsilon/2} \frac{\sigma_1}{g} \frac{  2 \zeta' \tau + \chi' Q }{2 \zeta'}    + \mu^{-\epsilon/2} \frac{\sigma_2}{g}  \frac{ Q}{2 \zeta'  }  +\frac{1}{2} \mu^{-\epsilon/2} \frac{\sigma_3 -\ii  \sigma_4}{g} \overline G+ \frac{1}{2} \frac{\sigma_3 +\ii \sigma_4}{g} \mu^{-\epsilon/2}  G  \right)  \right]}\;,
\end{multline}
with $\phi = ( A_\mu, c,\overline c, b,\varphi,\overline \varphi,\omega ,\overline \omega )$ and
\begin{align}
&\int \d^d x \mathcal L(\phi, \sigma_1,\ldots ,\sigma_4)  =S_\GZ +  \int \d^d x \Biggl(\frac{1}{4 \zeta^{\prime } } \frac{\sigma_1^2}{g^2} +    \frac{b}{2 \zeta'   } \frac{\sigma_1 }{g} \mu^{\epsilon/2} A^2   + \frac{b^2  }{4 \zeta' } \mu^{\epsilon} (A_\mu^a A_\mu^a)^2   \nonumber\\
& + \frac{1}{4 \zeta' [4 \alpha' \zeta' - \chi^{\prime 2}  ] } \frac{\sigma_2^2}{g^2}  + \frac{b \chi'  - 2 a \zeta'}{2 \zeta' [4 \alpha' \zeta' - \chi^{\prime 2}  ] }  \mu^{\epsilon/2} \frac{\sigma_2 }{g}   A^2 -    \frac{c }{  4 \alpha' \zeta' - \chi^{\prime 2}   }   \mu^{\epsilon/2}  \frac{\sigma_2}{g} \overline \varphi \varphi  \nonumber \\
& + \frac{(b \chi' - 2 a \zeta')^2}{4 \zeta' [4 \alpha' \zeta' - \chi^{\prime 2}  ] }  \mu^{\epsilon} (A_\mu^a A_\mu^a)^2   +  \frac{c^2 \zeta' }{ [4 \alpha' \zeta' - \chi^{\prime 2}  ] }  \mu^{\epsilon} (\overline \varphi^a_i \varphi^a_i)^2  -  \frac{c (b \chi' - 2 a \zeta') }{ 4 \alpha' \zeta'  - \chi^{\prime 2}   }  \mu^{\epsilon}  A_\mu^a A_\mu^a  \overline\varphi^b_i \varphi^b_i  \nonumber\\
&   + \frac{1}{ 4 Z_\varrho Z_G^2  \varrho }\left( \frac{\sigma_3^2}{g^2} + \frac{\sigma_4^2}{g^2} \right)  + \mu^{\epsilon/2}\frac{Z_\varphi}{ 4 Z_\varrho  Z_{G} \varrho } \frac{\sigma_3 }{g} \left( \overline \varphi \overline \varphi + \varphi  \varphi \right)   +\mu^{\epsilon/2}\frac{Z_\varphi}{ 4 Z_\varrho  Z_{ G} \varrho } \frac{\ii \sigma_4 }{g}\left( \overline \varphi \overline \varphi - \varphi  \varphi \right) \nonumber\\
&+ \mu^\epsilon \frac{Z_\varphi^2}{4 Z_\varrho \varrho} \overline \varphi^a_i \overline \varphi^a_i \varphi^b_j \varphi^b_j \Biggr) \;.
\end{align}

\noindent As these HS transformations do not put everything in the right form yet, we propose the following \textit{extra} transformation
\begin{eqnarray}
\sigma_1 \frac{\chi'}{2 \zeta'} - \frac{\sigma_2}{2 \zeta'} &=& \sigma_2' \;.
\end{eqnarray}
So \eqref{enfunc2} becomes
\begin{multline}
\e^{- W(Q, \tau, G, \overline G)} = \int [\d A_\mu][\d c][\d\overline c][\d b] [\d \sigma_1][\d \sigma_2][\d \sigma_3] [\d \sigma_4] [\d \varphi] [\d\overline \varphi] [\d \omega] [\d \overline \omega] \\
\times \e^{\left[ - \int\d^d x \left( \mathcal L(\phi, \sigma_1, \ldots,\sigma_4)   - \mu^{-\epsilon/2} \frac{\sigma_1}{g}  \tau     - \mu^{-\epsilon/2} \frac{\sigma_2'}{g}   Q   +\frac{1}{2} \mu^{-\epsilon/2} \frac{\sigma_3 -\ii  \sigma_4}{g} \overline G+ \frac{1}{2} \frac{\sigma_3 +\ii \sigma_4}{g} \mu^{-\epsilon/2}  G  \right)  \right]}\;,
\end{multline}
whereby
\begin{align}\label{lagrangian}
&\int \d^d x \mathcal L(\phi, \sigma_1,\ldots ,\sigma_4)  =S_\GZ +  \int \d^d x \Biggl(  \frac{ \alpha' }{4 \alpha' \zeta' - \chi^{\prime 2}}  \frac{\sigma_1^2}{g^2} + \frac{ \zeta' }{ 4 \alpha' \zeta' - \chi^{\prime 2}}  \frac{\sigma_2^2}{g^2}   -  \frac{ \chi' }{4 \alpha' \zeta' - \chi^{\prime 2}}  \frac{\sigma_1 \sigma_2}{g^2}  \nonumber\\
 & +   \frac{2 b \alpha' - a \chi'}{ 4 \alpha' \zeta' - \chi^{\prime 2}  } \frac{\sigma_1 }{g} \mu^{\epsilon/2} A^2  -    \frac{b \chi'  - 2 a \zeta'}{ [4 \alpha' \zeta' - \chi^{\prime 2}  ] }  \mu^{\epsilon/2} \frac{\sigma_2 }{g}   A^2  -    \frac{c \chi' }{  4 \alpha' \zeta' - \chi^{\prime 2}   }   \mu^{\epsilon/2}  \frac{\sigma_1}{g} \overline \varphi \varphi \nonumber\\
&  +    \frac{ 2 c \zeta' }{  4 \alpha' \zeta' - \chi^{\prime 2}   }   \mu^{\epsilon/2}  \frac{\sigma_2}{g} \overline \varphi \varphi  + \frac{b^2  }{4 \zeta' } \mu^{\epsilon} (A_\mu^a A_\mu^a)^2  + \frac{(b \chi' - 2 a \zeta')^2}{4 \zeta' [4 \alpha' \zeta' - \chi^{\prime 2}  ] }  \mu^{\epsilon} (A_\mu^a A_\mu^a)^2    \nonumber\\
&  +  \frac{c^2 \zeta' }{ [4 \alpha' \zeta' - \chi^{\prime 2}  ] }  \mu^{\epsilon} (\overline \varphi^a_i \varphi^a_i)^2  -  \frac{c (b \chi' - 2 a \zeta') }{ 4 \alpha' \zeta'  - \chi^{\prime 2}   }   \mu^{\epsilon}  A_\mu^a A_\mu^a  \overline\varphi^b_i \varphi^b_i  + \frac{1}{ 4 Z_\varrho Z_G^2  \varrho }\left( \frac{\sigma_3^2}{g^2}+ \frac{\sigma_4^2}{g^2} \right) \nonumber\\
 & + \mu^{\epsilon/2}\frac{Z_\varphi}{ 4 Z_\varrho  Z_{G} \varrho } \frac{\sigma_3 }{g} \left( \overline \varphi \overline \varphi + \varphi  \varphi \right)   +\mu^{\epsilon/2}\frac{Z_\varphi}{ 4 Z_\varrho  Z_{ G} \varrho } \frac{\ii \sigma_4 }{g}\left( \overline \varphi \overline \varphi - \varphi  \varphi \right) + \mu^\epsilon \frac{Z_\varphi^2}{4 Z_\varrho \varrho} \overline \varphi^a_i \overline \varphi^a_i \varphi^b_j \varphi^b_j \Biggr)\;.
\end{align}
Now acting with $\left . \frac{\delta }{\delta Q}\right|_{Q,\tau = 0}$ and $\left. \frac{\delta}{\delta \tau} \right|_{Q,\tau =0}$ on the  energy functional, before and after the HS transformation,  gives us the following two relations,
\begin{eqnarray}
Z_{QQ} Z_\varphi\Braket{\overline \varphi^a_i \varphi^a_i  } + \frac{1}{2} Z_A Z_{\tau W } \Braket {A_{\mu }^{a}A_{\mu }^{a}}  &=& - \mu^{-\epsilon/2}
\frac{\Braket{\sigma_2}}{g}  \;, \nonumber\\
\frac{1}{2} Z_A Z_{\tau \tau } \Braket {A_{\mu }^{a}A_{\mu }^{a}}  &=& - \mu^{-\epsilon/2}
\frac{\Braket{\sigma_1}}{g} \;,
\end{eqnarray}
while acting with $\left . \frac{\delta }{\delta G}\right|_{G,\overline G = 0}$ and $\left. \frac{\delta}{\delta \overline G} \right|_{G,\overline G=0}$
\begin{eqnarray}
Z_G Z_\varphi \Braket{\varphi \varphi} &=& \mu^{-\epsilon/2} \frac{\Braket{\sigma_3 + \ii \sigma_4}}{g} \;, \nonumber\\
Z_{ G} Z_\varphi \Braket{\overline \varphi \overline \varphi} &=&  \mu^{-\epsilon/2}\frac{\Braket{\sigma_3 -\ii \sigma_4}}{g} \;,
\end{eqnarray}
or equivalently
\begin{eqnarray}
Z_G Z_\varphi  \frac{1}{2}\Braket{\varphi \varphi + \overline \varphi \overline \varphi} &=& \mu^{-\epsilon/2} \frac{\Braket{\sigma_3}}{g} \;, \nonumber\\
Z_G Z_\varphi  \frac{\ii}{2}\Braket{ \overline \varphi \overline \varphi- \varphi \varphi } &=& \mu^{-\epsilon/2} \frac{\Braket{\sigma_4}}{g}\;.
\end{eqnarray}

\subsection{The effective action}
If we  introduce the parameters
\begin{eqnarray}
\frac{m^2}{2} &=&   \frac{1}{4\alpha_0 \zeta_0 - 2 \chi_0^2} \left( 2 \alpha_0  g \sigma_1 - \chi_0 g \sigma_2\right) \;, \nonumber\\
M^2 &=&   \frac{1}{2\alpha_0 \zeta_0 -  \chi_0^2} \left(  \chi_0  g \sigma_1 - \zeta_0 g \sigma_2\right) \;, \nonumber\\
\rho &=& -\frac{53 N}{ 48 (N^2-1)^2   }  (\sigma_3 + \ii \sigma_4) g  \;, \nonumber\\
\rho^\dagger &=& -\frac{53 N}{ 48 (N^2-1)^2   }  (\sigma_3 - \ii \sigma_4)g \;,
\end{eqnarray}
with $\alpha_0, \zeta_0, \chi_0$ given in equations \eqref{weljui}-\eqref{opl2}, then the quadratical part of the Lagrangian \eqref{lagrangian} is given by
\begin{align}
&\int \d^d x \mathcal L(\phi, \sigma_1,\ldots ,\sigma_4)  =S_\GZ^{\quadr} +  \int \d^d x \Biggl(   \frac{ \alpha' }{4 \alpha' \zeta' - \chi^{\prime 2}}  \frac{\sigma_1^2}{g^2} + \frac{ \zeta' }{ 4 \alpha' \zeta' - \chi^{\prime 2}}  \frac{\sigma_2^2}{g^2}   -  \frac{ \chi' }{4 \alpha' \zeta' - \chi^{\prime 2}}  \frac{\sigma_1 \sigma_2}{g^2} \nonumber\\
 & + \frac{1}{ 4 Z_\varrho Z_G^2  \varrho }\left( \frac{\sigma_3^2}{g^2} + \frac{\sigma_4^2}{g^2} \right)
  +  \frac{ m^2}{2} \mu^{\epsilon/2} A^2  -    M^2 \mu^{\epsilon/2}  \overline \varphi \varphi     + \mu^{\epsilon/2} \frac{\rho}{2} \overline \varphi \overline \varphi +  \mu^{\epsilon/2} \frac{\rho^\dagger}{2} \varphi  \varphi  \Biggr)\;.
\end{align}
We have left out the higher order terms as we shall only calculate the one loop effective potential  $\Gamma^{(1)}$ . \\
\\
 All details of the calculations of the effective potential have been collected in the  appendix \ref{detailsvery}.   The final result for the effective potential $\Gamma^{(1)}$ is given by
\begin{align}\label{finaleffective}
\Gamma^{(1)} &=  \frac{(N^2 -1)^2}{16 \pi^2} \Bigl[(M^2 - \sqrt{\rho \rho^\dagger})^2 \ln\frac{M^2 - \sqrt{\rho \rho^\dagger}}{\overline \mu^2}  + (M^2 + \sqrt{\rho \rho^\dagger})^2 \ln\frac{M^2 + \sqrt{\rho \rho^\dagger}}{\overline \mu^2} \nonumber\\
& - 2 (M^2 + \rho \rho^\dagger)\Bigr] + \frac{3(N^2 -1)}{64\pi^2} \Bigl[  - \frac{5}{6} (m^4 -2 \lambda^4) + y_1^2 \ln \frac{(-y_1)}{\overline \mu} + y_2^2 \ln \frac{(-y_2)}{\overline \mu} + y_3^2 \ln \frac{(-y_3)}{\overline \mu} \nonumber\\
&- y_4^2 \ln \frac{(-y_4)}{\overline \mu} - y_5^2 \ln \frac{(-y_5)}{\overline \mu} \Bigr]  -2 (N^2 - 1) \frac{\lambda^4}{N g^2}  + \frac{3}{2} \frac{\lambda^4}{32 \pi^2} (N^2-1) \nonumber\\
&+ \frac{1}{2} \frac{48 (N^2-1)^2}{53 N} \left( 1   - N g^2 \frac{53}{24} \frac{\varrho_1}{(N^2 - 1)^2} \right) \frac{\rho \rho^\dagger}{g^2} \nonumber\\
&  +  \frac{9}{13} \frac{N^2-1}{N}\frac{m^4}{2g^2}- \frac{24}{35}\frac{(N^2-1)^2}{N}\frac{M^4}{g^2}  - \frac{161}{52} \frac{N^2-1}{16 \pi^2}\frac{ m^4}{2 }- M^4 \alpha_1  +M^2 m^2 \chi_1\;.
\end{align}
whereby $y_1$, $y_2$ and $y_3$ are the solutions of the equation $y^3+(m^2 +2 M^2) y^2 +\bigl(\lambda^4+ M^4- \rho \rho^\dagger +2 M^2 m^2 \bigr) y + M^2 \lambda^4 + 1/2 ( \rho + \rho^\dagger) \lambda^4 + M^4 m^2  - m^2 \rho \rho^\dagger  =0$ and $y_4$ and $y_5$ of the equation $ y^2 + 2 M^2 y +M^4 -\rho \rho^\dagger =0$. We employed the $\MSbar$ scheme.

\subsection{Minimizing the effective potential to prove that the condensates are non-vanishing}
To simplify the calculations, let us set $\rho= \rho^\dagger = 0$, which  corresponds to the case of not considering the condensates $\braket{\overline \varphi \overline \varphi}$ and $\Braket{\varphi \varphi}$. For  the  moment, we are only considering $\braket{\varphi \overline \varphi}$, which already has the desired influence on the propagators, see the next section. In this case, the effective action  simplifies, and becomes:
\begin{align}\label{mineffpot}
\Gamma^{(1)} &=  \frac{(N^2 -1)^2}{16 \pi^2}\Bigl[2 M^4 \ln\frac{M^2}{\overline \mu^2}   - 2 M^2 \Bigr] + \frac{3(N^2 -1)}{64\pi^2} \Bigl[  - \frac{5}{6} (m^4 -2 \lambda^4)+ M^4 \ln \frac{(M^2)}{\overline \mu} \nonumber\\
&+ y_2^2 \ln \frac{(-y_2)}{\overline \mu} + y_3^2 \ln \frac{(-y_3)}{\overline \mu} - 2M^4 \ln \frac{M^2}{\overline \mu}  \Bigr]  -2 (N^2 - 1) \frac{\lambda^4}{N g^2}  + \frac{3}{2} \frac{\lambda^4}{32 \pi^2} (N^2-1) \nonumber\\
&  +  \frac{9}{13} \frac{N^2-1}{N}\frac{m^4}{2g^2}- \frac{24}{35}\frac{(N^2-1)^2}{N}\frac{M^4}{g^2}  - \frac{161}{52} \frac{N^2-1}{16 \pi^2}\frac{ m^4}{2 }- M^4 \alpha_1  +M^2 m^2 \chi_1\;.
\end{align}
whereby $y_2$ and $y_3$ are are given by $\frac{1}{2} \left(-m^2-M^2\pm\sqrt{m^4-2 M^2 m^2 + M^4 - 4 \lambda^4}\right)$. \\
\\
In order to find the minimum, we should derive this action w.r.t.~$m^2$ and $M^2$ and  set the equations equal to zero. In addition, we should also impose the horizon condition \eqref{gapgamma}. Therefore, we have the following three conditions,
\begin{align}
\frac{\p \Gamma}{\p M^2} &=0 \;, &  \frac{\p \Gamma}{\p m^2} &=0 \;, &  \frac{\p \Gamma}{\p \lambda^4} &=0 \;,
\end{align}
which have to be solved for $M^2$, $m^2$ and $\lambda^4$. Unfortunately, it is impossible to solve these equations exactly due to the two unknown parameters $\alpha_1$ and $\chi_1$. However, we would like to know if the condensate $\braket{\overline \varphi \varphi}$ is present or not. For this, we need to  uncover if $M^2 = 0$ can be a solution of the above expression.  We can strongly argue that this is not the case, and thus that $M^2 \not=0$. \\
\\
We shall start from expression \eqref{mineffpot} and derive w.r.t.~$M^2$, $m^2$ and $\lambda^4$. As we would like to know if $M^2 = 0$ can be a minimum of the potential, we further set $M^2 = 0$. We then obtain the following equations
\begin{align}
&\frac{3 \left(\ln \left(m^2-\sqrt{m^4-4 \lambda^4}\right)-\ln \left(m^2+\sqrt{m^4-4 \lambda^4}\right)\right) \lambda^4}{4 \pi ^2 \sqrt{m^4-4 \lambda^4}}+m^2 \chi_1 -\frac{8}{\pi ^2}  = 0 \;, \nonumber\\
& \frac{1}{ \sqrt{m^4-4 \lambda^4}} \Biggl[ 11 \sqrt{m^4 - 4 \lambda^4} (24 \ln 2 -17 ) m^2\nonumber\\
 & +39 \left(-m^4+\sqrt{m^4-4 \lambda^4} m^2+2 \lambda^4\right) \ln \left(\frac{1}{8} \left(m^2-\sqrt{m^4-4 \lambda^4}\right)\right)\nonumber\\
&  +39 \left(m^4+\sqrt{m^4-4 \lambda^4} m^2-2 \lambda^4\right) \ln \left(\frac{1}{8} \left(m^2+\sqrt{m^4-4 \lambda^4}\right)\right) \Biggr] = 0 \;, \nonumber\\
&\frac{ \lambda^2 }{ \sqrt{m^4-4 \lambda^4}} \Biggl[ 9 \left(\sqrt{m^4-4 \lambda^4}-m^2\right)\ln \left(\frac{1}{8} \left(m^2-\sqrt{m^4-4 \lambda^4}\right)\right)+9 \left(m^2+\sqrt{m^4-4 \lambda^4}\right)\nonumber\\
& \times  \ln \left(\frac{1}{8} \left(m^2+\sqrt{m^4-4 \lambda^4}\right)\right)+\sqrt{m^4-4 \lambda^4} (-15+176 \ln 2 ) \Biggr] = 0\;,
\end{align}
whereby we have chosen to set\footnote{We work in units $\lms=1$.} $\overline \mu = 2  $ and $N = 3$. Now looking at the equation, we see that the second and third equation can be solved exactly for $m^2$ and $\lambda$. There are even multiple solutions possible. We take the solution which has the lowest value for the effective action with $M^2 = 0$. However, for this solution to be also a solution of the first equation, these values should be very specific and the chance that they will also satisfy the first equation is practically non-existent, with a certain value of $\chi_1$. Moreover, at a different scale $\overline \mu$, the three equations will look slightly different. However, $\chi_1$ is a number and stays the same. Therefore, it would be necessary  that at all different scales  these three equations can be solved exactly for only two parameters. This is practically impossible, leading to the conclusion that $M^2 \not= 0$. A similar reasoning can be worked out if $\rho$ and/or $\rho^\dagger$ would be allowed. The main result is that it is impossible for all these condensates to be zero, making the associated refinement inevitable.\\
\\
In conclusion, we have a firm indication that the condensate $\braket{\overline \varphi \varphi}$ is indeed present, thereby suggesting the dynamical transformation of the GZ framework into a refined GZ framework, with associated propagators that are in agreement with  the most recent  lattice data of \cite{Cucchieri:2007md,Bogolubsky:2007ud,Cucchieri:2007rg,Dudal:2010tf}.

\section{The gluon and the ghost propagator}
\subsection{The gluon propagator}
The gluon propagator shall still be infrared suppressed and non-zero at zero momentum. Indeed, starting from the further refined action \eqref{CGZ}, the quadratic action is given by
\begin{eqnarray}
S_{\quadr} &=& \frac{1}{4} (\p_\mu A_\nu - \p_\nu A_\mu)^2 + b \p_\mu A_\mu + \overline c \p^2 c + \overline \varphi \p^2 \varphi - \overline \omega \p^2 \omega - \gamma^2 g f^{abc} A_\mu^b (\varphi^{bc}_\mu + \overline \varphi^{bc}_\mu )\nonumber\\
&&  + \gamma^4 d (N^2 - 1) - M^2 \overline \varphi \varphi + \frac{m^2}{2} A_\mu A_\mu - \frac{\rho}{2} \overline \varphi \overline \varphi -\frac{\rho^{\dagger}}{2}\varphi  \varphi \;,
\end{eqnarray}
whereby we have replaced the source $\tau$ with $m^2$, $Q$ with $-M^2$, $\overline G^{ij}$ with $-\delta^{ij} \rho$ and $G^{ij}$ with $-\delta^{ij} \rho^{\dagger}$ and set all other sources equal to zero. From this, we can easily deduce the gluon propagator
\begin{multline} \label{gluonpropvra}
 \Braket{ A^a_{\mu}(p) A^b_{\nu}(-p)}  =  \left[\delta_{\mu\nu} - \frac{p_{\mu}p_{\nu}}{p^2} \right]\delta^{ab} \\\underbrace{ \frac{2 \left(M^2+p^2\right)^2-2 \rho  \rho^{\dagger}  }{ 2 M^4 p^2+2 p^6+2 M^2 \left(2 p^4+\lambda^4 \right)-\lambda^4  (\rho +\rho^{\dagger} )+2 m^2 \left(\left(M^2+p^2\right)^2-\rho  \rho^{\dagger} \right)+2 p^2 (\lambda^4 -\rho  \rho^{\dagger} ) }}_{\mathcal{D}(p^2)}\;,
\end{multline}
with $\lambda^4 = 2 g^2 N \gamma^4$. If we assume that $\rho = \rho^\dagger$, we then find the following gluon propagator:
\begin{eqnarray}
D(p^2) = \frac{M^2+p^2+\rho }{p^4 + M^2 p^2 +p^2 (\rho + m^2) +m^2 \left(M^2+\rho \right) +  \lambda ^4} \;,
\end{eqnarray}
which has exactly the same form as the refined gluon propagator \eqref{gluonprop2}. However, for the moment we cannot say whether $\rho = \rho^\dagger$ is the case or not. This shall be further investigated in \cite{tobewritten} upon using lattice input. Notice that $\rho$, $\rho^\dagger$ as well as $M^2$  provide in an independent way that $D(0) \not= 0$. In principle, it could occur that $M^4 = \rho \rho^\dagger$, giving $D(0) = 0$, but there is no obvious reason why this relation should have to hold.

\subsection{The ghost propagator}
The one loop ghost propagator is given by
\begin{eqnarray} \label{ghostpropagator1}
\mathcal{G}^{ab}(k^2) &=&  \delta^{ab} \mathcal{G}(k^2) ~=~ \delta^{ab}\left( \frac{1}{k^2} +
\frac{1}{k^2} \left[g^2 \frac{N}{N^2 - 1} \int \frac{\d^4
q}{(2\pi)^4} \frac{(k-q)_{\mu} k_{\nu}}{(k-q)^2}
 \Braket{A^a_{\mu}A^a_{\nu}}\right] \frac{1}{k^2} \right) + \mathcal{O}(g^4) \nonumber\\
&=& \delta^{ab} \frac{1}{k^2} (1+ \sigma(k^2)) +
\mathcal{O}(g^4) =  \delta^{ab} \frac{1}{k^2 (1 - \sigma(k^2))} +
\mathcal{O}(g^4) \;,
\end{eqnarray}
with
\begin{align*}
&\sigma(k^2)=\frac{N}{N^2 - 1} \frac{g^2}{k^2}\int \frac{\d^4 q}{(2\pi)^4} \frac{(k-q)_{\mu} k_{\nu}}{(k-q)^2} \Braket{A^a_{\mu}A^a_{\nu}} \nonumber\\
&= Ng^2 \frac{k_{\mu} k_{\nu}}{k^2} \int \frac{\d^d q}{(2\pi)^d} \frac{1}{(k-q)^2} \left[ \delta_{\mu\nu} - \frac{q_{\mu}q_{\nu}}{q^2} \right]\nonumber\\
 &\times\frac{2 \left(M^2+q^2\right)^2-2 \rho  \rho^{\dagger}  }{ 2 M^4 q^2+2 q^6+2 M^2 \left(2 q^4+\lambda^4 \right)-\lambda^4  (\rho +\rho^{\dagger} )+2 m^2 \left(\left(M^2+q^2\right)^2-\rho  \rho^{\dagger} \right)+2 q^2 (\lambda^4 -\rho  \rho^{\dagger} ) } \;.
\end{align*}
As we are interested in the infrared behavior of this propagator, we expand the previous expression for small $k^2$
\begin{multline}\label{sigmaex}
\sigma (k^2\approx 0) =   Ng^2 \frac{d-1}{d}\int \frac{\d^d q}{(2\pi)^d}  \frac{1}{q^2} \\
\times \frac{2 \left(M^2+q^2\right)^2-2 \rho  \rho^{\dagger}  }{ 2 M^4 q^2+2 q^6+2 M^2 \left(2 q^4+\lambda^4 \right)-\lambda^4  (\rho +\rho^{\dagger} )+2 m^2 \left(\left(M^2+q^2\right)^2-\rho  \rho^{\dagger} \right)+2 q^2 (\lambda^4 -\rho  \rho^{\dagger} ) } \\ + O(k^2) \;.
\end{multline}
Let us now have a look at the gap equation. For this we can start from the (one-loop) effective action which can be written as (see the appendix \ref{detailsvery})
\begin{equation*}
\Gamma_\gamma^{(1)} = -d(N^{2}-1)\gamma^{4} +\frac{(N^{2}-1)}{2}\left( d-1\right) \int \frac{\d^{d}q}{\left( 2\pi \right) ^{d}} \ln A  + \ldots\;,
\end{equation*}
with \begin{equation*}
 A= \frac{2 M^4 q^2+2 q^6+2 M^2 \left(2 q^4+\lambda^4 \right)-\lambda^4  (\rho +\rho^{\dagger} )+2 m^2 \left(\left(M^2+q^2\right)^2-\rho  \rho^{\dagger} \right)+2 q^2 (\lambda^4 -\rho  \rho^{\dagger} )}{2 \left(M^2+q^2\right)^2-2 \rho  \rho^{\dagger}  } \;,
\end{equation*}
and the $\ldots$ indicating parts independent from $\lambda$. Setting $\lambda^4 = 2 g^2 N \gamma^4$, we rewrite the previous expression,
\begin{eqnarray*}
\mathcal{E}^{(1)} &=&  \frac{\Gamma_\gamma^{(1)}}{N^2 - 1} \frac{2 g^2 N}{d} ~=~ - \lambda^4  + g^2 N \frac{ d-1}{d} \int \frac{\d^{d}q}{\left(2\pi \right) ^{d}} \ln A + \ldots \;.
\end{eqnarray*}
The gap equation is given by $\frac{\p \mathcal{E}^{(1)}  }{\p \lambda^2} = 0$,
\begin{multline}
1 =    g^2 N \frac{d-1}{d} \int \frac{\d^{d}q}{\left(
2\pi \right) ^{d}}  \\
 \frac{ 2 M^2+2 q^2 - \rho -\rho^{\dagger} }{2 M^4 q^2+2 q^6+2 M^2 \left(2 q^4+\lambda^4 \right)-\lambda^4  (\rho +\rho^{\dagger} )+2 m^2 \left(\left(M^2+q^2\right)^2-\rho  \rho^{\dagger} \right)+2 q^2 (\lambda^4 -\rho  \rho^{\dagger} ) } ,
\end{multline}
where we have excluded the solution $\lambda =0$. With the help of this gap equation, we can rewrite equation \eqref{sigmaex},
\begin{multline}\label{sigmaex1}
\sigma (k^2\approx 0) = 1 +   Ng^2 \frac{d-1}{d}\int \frac{\d^d q}{(2\pi)^d}  \\
\times \frac{ 2 M^4/ q^2 + 2 M^2 - 2 \rho \rho^\dagger / q^2 + \rho + \rho^\dagger  }{ 2 M^4 q^2+2 q^6+2 M^2 \left(2 q^4+\lambda^4 \right)-\lambda^4  (\rho +\rho^{\dagger} )+2 m^2 \left(\left(M^2+q^2\right)^2-\rho  \rho^{\dagger} \right)+2 q^2 (\lambda^4 -\rho  \rho^{\dagger} ) } \\
+ O(k^2) \;.
\end{multline}
The integral in the above expression is finite. We can rewrite the integral as ($d = 4$)
\begin{multline*}
 I= Ng^2 \frac{3}{32 \pi^2}\int_0^\infty \d q \\
  \frac{ q ( M^4 -  (r^2 + s^2) )  +  q^3(  M^2  + r)  }{  M^4 q^2+  q^6+ M^2 \left(2 q^4+\lambda^4 \right)- r \lambda^4  +  m^2 \left(\left(M^2+q^2\right)^2-( r^2 + s^2)  \right)+ q^2 (\lambda^4 -(r^2 + s^2) ) }\;,
\end{multline*}
with $I = \sigma (k^2\approx 0)  - 1$, whereby we have parameterized
\begin{align}
\rho &= r + \ii s \;,  & \rho^\dagger &= r - \ii s \;.
\end{align}
We further write
\begin{multline}
 I =    \frac{3 Ng^2}{64 \pi^2} \int_0^\infty \d x \left(  M^4 -  (r^2 + s^2)   +  x (  M^2  + r)  \right) / \left(  x^3 + x^2 (2M^2 + m^2 ) \right. \\
  \left.  + x (M^4 + 2 m^2 M^2 +  \lambda^4 - (r^2 + s^2)) + \lambda^4 (M^2 - r) + m^2 (M^4 - (r^2 + s^2)) \right)\;.
\end{multline}
\\
\textbf{Solution of cubic equation}\\
\\
The next step would be to solve the cubic equation in the  denominator  of the equation above,
\begin{equation}
 x^3 + x^2 \underbrace{(2M^2 + m^2 )}_a + x \underbrace{(M^4 + 2 m^2 M^2 + \lambda^4 - (r^2 + s^2))}_b   + \underbrace{\lambda^4 (M^2 - r) + m^2 (M^4 - (r^2 + s^2)) }_c = 0 \;.
\end{equation}
In general, the  roots  are given by
\begin{eqnarray}
x_1 &=& \frac{-1}{3} \left( a  + \sqrt[3]{ \frac{m + \sqrt{n}}{2} } + \sqrt[3]{ \frac{m - \sqrt{n}}{2} } \right) \;, \nonumber\\
x_2 &=& \frac{-1}{3} \left( a  + \frac{-1 + \ii \sqrt{3}}{2} \sqrt[3]{ \frac{m + \sqrt{n}}{2} } + \frac{-1 - \ii \sqrt{3}}{2} \sqrt[3]{ \frac{m - \sqrt{n}}{2} } \right) \;, \nonumber\\
x_3 &=& \frac{-1}{3} \left( a  + \frac{-1 - \ii \sqrt{3}}{2} \sqrt[3]{ \frac{m + \sqrt{n}}{2} } + \frac{-1 + \ii \sqrt{3}}{2} \sqrt[3]{ \frac{m - \sqrt{n}}{2} } \right)\;,
\end{eqnarray}
with
\begin{eqnarray}
m &=& 2 \left(m^2-M^2\right) \left(\left(m^2-M^2\right)^2-9 \left(r^2+s^2\right)\right)-9 \left(m^2-M^2+3 r\right) \lambda ^4 \;,\nonumber\\
n&=& \left[2 \left(m^2-M^2\right) \left(\left(m^2-M^2\right)^2-9 \left(r^2+s^2\right)\right)-9 \left(m^2-M^2+3 r\right) \lambda ^4\right]^2 \nonumber\\
 &&-4 \left[\left(m^2-M^2\right)^2+3 \left(r^2+s^2-\lambda ^4\right)\right]^3\;.
\end{eqnarray}
Of course, it is possible that two (or three) solutions coincide. This can be checked by calculating the discriminant
\begin{equation}
\Delta = - 4 a^3 c + a^2 b^2 - 4 b^3 + 18 abc - 27 c^2 \;.
\end{equation}
If $\Delta =0$,  then the equation has three  real roots and at least two are equal. \\
\\
\textbf{Case 1: $x_1 \not = x_2 \not = x_3$}\\
\\
If $x_1 \not = x_2 \not = x_3$, we can rewrite the integral  $I$ as,
\begin{multline}
I=Ng^2 \frac{3}{64 \pi^2}\Biggl[  \int_0^\infty \d x  \frac{ M^4 -  (r^2 + s^2) + (  M^2  + r)x_1  }{(x_1-x_2)(x_1-x_3)} \frac{1}{x-x_1} \\
 + \int_0^\infty \d x  \frac{ M^4 -  (r^2 + s^2) + (  M^2  + r)x_2  }{(x_2-x_3)(x_2-x_1)(x-x_2)}+ \int_0^\infty \d x  \frac{ M^4 -  (r^2 + s^2) + (  M^2  + r)x_3  }{(x_3-x_1)(x_3-x_1)(x-x_3)} \Biggr]\;.
\end{multline}
These integrals are now easy to solve,  they all are of the type $\int \d x \frac{1}{x} = \ln x$.
\begin{multline*}
I=Ng^2 \frac{3}{64 \pi^2}\Biggl[ \underbrace{ \frac{ M^4 -  (r^2 + s^2) + (  M^2  + r)x_1  }{(x_1-x_2)(x_1-x_3)} }_{u_1} \left. \ln (x - x_1 )  \right|^{\infty}_0 \\
+   \underbrace{\frac{ M^4 -  (r^2 + s^2) + (  M^2  + r)x_2  }{(x_2-x_3)(x_2-x_1)} }_{v_1} \left. \ln (x - x_2 )  \right|^{\infty}_0  + \underbrace{ \frac{ M^4 -  (r^2 + s^2) + (  M^2  + r)x_3  }{(x_3-x_1)(x_3-x_1)}}_{w_1}  \left. \ln (x - x_3 )  \right|^{\infty}_0 \Biggr]\;.
\end{multline*}
One could expect there is a problem at infinity, in contrast with what we have concluded before. However, as $u_1 + v_1 + w_1 = 0$, the infinities cancel. We obtain,
\begin{multline}
I=Ng^2 \frac{3}{64 \pi^2}\Biggl[  \frac{ M^4 -  (r^2 + s^2) + (  M^2  + r)x_1  }{(x_1-x_2)(x_1-x_3)}   \ln ( - x_1 ) \\
 +   \frac{ M^4 -  (r^2 + s^2) + (M^2 + r)x_2  }{(x_2-x_3)(x_2-x_1)}\ln ( - x_2 ) +  \frac{ M^4 -  (r^2 + s^2) + (  M^2  + r)x_3  }{(x_3-x_1)(x_3-x_1)} \ln (- x_3 )  \Biggr]\;.
\end{multline}
\\
\textbf{Case 2: $x_1  = x_2 \not = x_3$}\\
\\
In this case, we can rewrite the integral   $I$ as
\begin{eqnarray}
I&=& Ng^2 \frac{3}{64 \pi^2}\Biggl[  \int_0^\infty \d x \underbrace{ \frac{ M^4 -  (r^2 + s^2) + (  M^2  + r)x_1  }{ (x_1-x_3)^2}}_{u_2} \frac{1}{x-x1}  \nonumber\\
 && - \int_0^\infty \d x  \underbrace{\frac{ M^4 -  (r^2 + s^2) + (  M^2  + r)x_1  }{ (x_1-x_3)^2}}_{v_2} \frac{1}{x-x_3} \nonumber\\
 && + \int_0^\infty \d x  \underbrace{\frac{ M^4 -  (r^2 + s^2) + (  M^2  + r)x_3  }{x_1-x_3}}_{w_2} \frac{1}{(x-x_3)^2} \Biggr]\;.
\end{eqnarray}
One can check that $u_2 + v_2 =0$, so we can  perform the integrations,
\begin{multline}
I=Ng^2 \frac{3}{64 \pi^2}\Biggl[   \frac{ M^4 -  (r^2 + s^2) + (  M^2  + r)x_1  }{ (x_1-x_3)^2}  \ln (-x1)  \\
- \frac{ M^4 -  (r^2 + s^2) + (  M^2  + r)x_1  }{ (x_1-x_3)^2} \ln (-x_3)  -  \frac{ M^4 -  (r^2 + s^2) + (  M^2  + r)x_3  }{x_1-x_3}  \frac{1}{x_3^2} \Biggr]\;.
\end{multline}

\newpage

\textbf{Case 3: $x_1  = x_2  = x_3$}\\
\\
Finally, in this case we can write
\begin{equation}
I=Ng^2 \frac{3}{64 \pi^2}\Biggl[    ( M^2+ r )\int_0^\infty \d x   \frac{1}{ (x-x_1)^2 } + (M^4 -  (r^2 + s^2) + (  M^2  + r)x_1)\int_0^\infty \d x    \frac{1}{(x-x_1)^3} \Biggr]\;,
\end{equation}
so after integration
\begin{equation}
I=Ng^2 \frac{3}{64 \pi^2}\Biggl[   - \frac{ M^2+ r } { x_1 } + \frac{M^4 -  (r^2 + s^2) + (  M^2  + r)x_1}{ 2 x_1^2} \Biggr]\;.
\end{equation}
Now we can make some conclusions. Looking at the different cases, it looks almost certain that $I \not = 0$, as very specific values of the condensates would be needed to take care of this. Therefore, we have strong indications that the ghost propagator is  not enhanced, in addition to the nonvanishing gluon propagator at zero momentum.

\section{Conclusion}
Although this paper is quite technical, the conclusions are quite simple. Firstly, we have shown that using the GZ action, more condensates can influence the dynamics. We have  investigated  in detail the following condensates: $\braket{A_\mu^a A_\mu^a}, \braket{\overline \varphi^a_i \varphi^a_i}, \braket{\overline \varphi^a_i \overline \varphi^a_i}$ and $ \braket{ \varphi^a_i  \varphi^a_i}$ whereby the latter two were never investigated before. We have proven that we can renormalize the GZ action in the presence of these condensates. In particular, a   renormalizable effective potential, compatible with the renormalization group, can be constructed for the associated local composite operators.\\
\\
Secondly, for the first time, we were able to calculate the one loop effective potential in the LCO formalism:
\begin{align}\label{finaleffective}
\Gamma^{(1)} &=  \frac{(N^2 -1)^2}{16 \pi^2} \Bigl[(M^2 - \sqrt{\rho \rho^\dagger})^2 \ln\frac{M^2 - \sqrt{\rho \rho^\dagger}}{\overline \mu^2}  + (M^2 + \sqrt{\rho \rho^\dagger})^2 \ln\frac{M^2 + \sqrt{\rho \rho^\dagger}}{\overline \mu^2} \nonumber\\
& - 2 (M^2 + \rho \rho^\dagger)\Bigr] + \frac{3(N^2 -1)}{64\pi^2} \Bigl[  - \frac{5}{6} (m^4 -2 \lambda^4) + y_1^2 \ln \frac{(-y_1)}{\overline \mu} + y_2^2 \ln \frac{(-y_2)}{\overline \mu} + y_3^2 \ln \frac{(-y_3)}{\overline \mu} \nonumber\\
&- y_4^2 \ln \frac{(-y_4)}{\overline \mu} - y_5^2 \ln \frac{(-y_5)}{\overline \mu} \Bigr]  -2 (N^2 - 1) \frac{\lambda^4}{N g^2}  + \frac{3}{2} \frac{\lambda^4}{32 \pi^2} (N^2-1) \nonumber\\
&+ \frac{1}{2} \frac{48 (N^2-1)^2}{53 N} \left( 1   - N g^2 \frac{53}{24} \frac{\varrho_1}{(N^2 - 1)^2} \right) \frac{\rho \rho^\dagger}{g^2} \nonumber\\
&  +  \frac{9}{13} \frac{N^2-1}{N}\frac{m^4}{2g^2}- \frac{24}{35}\frac{(N^2-1)^2}{N}\frac{M^4}{g^2}  - \frac{161}{52} \frac{N^2-1}{16 \pi^2}\frac{ m^4}{2 }- M^4 \alpha_1  +M^2 m^2 \chi_1\;.
\end{align}
whereby $y_1$, $y_2$ and $y_3$ are the solutions of the equation $y^3+(m^2 +2 M^2) y^2 +\bigl(\lambda^4+ M^4- \rho \rho^\dagger +2 M^2 m^2 \bigr) y + M^2 \lambda^4 + 1/2 ( \rho + \rho^\dagger) \lambda^4 + M^4 m^2  - m^2 \rho \rho^\dagger  =0$ and $y_4$ and $y_5$ of the equation $ y^2 + 2 M^2 y +M^4 -\rho \rho^\dagger =0$. Unfortunately, due to the existence of yet unknown higher loop parameters, i.e.~$\alpha_1$, $\varrho_1$ and $\chi_1$,  in the one loop effective action, we are yet unable to give an estimate for the different condensates. Nevertheless, we have been able to already  provide strong indications that some condensates are in fact non-zero and shall lower the effective action. We hope to come back to the explicit computation of the parameters $\alpha_1$, $\varrho_1$ and $\chi_1$ in the future. In particular, one should compute the divergences of the vacuum diagram in Figure \ref{1loopd2}, the similar one for the mixing, and other divergent 2 loop diagrams stemming from the operators $\varphi \varphi$ and $\overline \varphi \overline \varphi$. Once this task will be executed, all information is available to actually work out the one loop effective potential and to investigate its structure and the associated formation of the RGZ condensates.\\
\\
Thirdly, we have also shown that in this further refined framework, the gluon propagator is non zero at zero momentum, and the ghost propagator will be non-enhanced.\\
\\
A complementary approach to the current one, is to find out to what extent a gluon propagator of the type \eqref{gluonpropvra} or ghost propagator of the type \eqref{ghostpropagator1} could describe the lattice data, not only qualitatively, but also quantitatively. This is current under investigation in \cite{tobewritten} for different space time dimensions. In \cite{Dudal:2010tf} it was already shown that a RGZ propagator \eqref{gluonpropvra} reproduces the SU(3) data very well. \\
\\
Another question which was not answered here, is whether $\sigma(k^2)$, see equation \eqref{ghostpropagator1}, is in fact smaller than one. This is necessary in order to be assured to stay  within the Gribov horizon. However, this question shall also be addressed in \cite{tobewritten}, and we refer to this paper for further details on this matter.

\section*{Acknowledgments}
D.~Dudal and N.~Vandersickel are supported by the Research-Foundation
Flanders (FWO Vlaanderen). S.~P.~Sorella is supported by the FAPERJ, Funda{%
\c c}{\~a}o de Amparo {\`a} Pesquisa do Estado do Rio de Janeiro, under the
program \textit{Cientista do Nosso Estado}, E-26/100.615/2007. The Conselho
Nacional de Desenvolvimento Cient\'{\i}fico e Tecnol\'{o}gico (CNPq-Brazil),
the Faperj, Funda{\c{c}}{\~{a}}o de Amparo {\`{a}} Pesquisa do Estado do Rio
de Janeiro, the SR2-UERJ and the Coordena{\c{c}}{\~{a}}o de Aperfei{\c{c}}%
oamento de Pessoal de N{\'{\i}}vel Superior (CAPES) are gratefully
acknowledged for financial support.

\newpage

\appendix

\section{Recapitulation of the Gribov-Zwanziger action and  of its renormalizability \label{appendixA}}
In this appendix, we shall repeat the complete  proof of the renormalization of the Gribov-Zwanziger action \cite{Dudal:2010fq}.

\subsection{The Gribov-Zwanziger action and the BRST  symmetry}
We start with the Gribov-Zwanziger action,
\begin{eqnarray}\label{GZstart}
S_\GZ &=& S_{0} +  S_{\gamma} \,,
\end{eqnarray}
with
\begin{eqnarray}
S_{0}&=&S_\YM + S_\gf + \int \d^d x \left( \overline \varphi_i^a \p_\mu \left( D_\mu^{ab} \varphi^b_i \right)  - \overline \omega_i^a \p_\mu \left( D_\mu^{ab} \omega_i^b \right) - g f^{abc} \p_\mu \overline \omega_i^a    D_\mu^{bd} c^d  \varphi_i^c \right) \nonumber \;, \\
S_{\gamma}&=& -\gamma ^{2}g\int\d^{d}x\left( f^{abc}A_{\mu }^{a}\varphi _{\mu }^{bc} +f^{abc}A_{\mu}^{a}\overline{\varphi }_{\mu }^{bc} + \frac{d}{g}\left(N^{2}-1\right) \gamma^{2} \right) \;.
\end{eqnarray}
We recall that we have simplified the notation of the additional fields $\left( \overline \varphi_\mu^{ac},\varphi_\mu^{ac},\overline \omega_\mu^{ac},\omega_\mu^{ac}\right) $ in $S_0$ as $S_0$ displays a symmetry with respect to the composite index $i=\left( \mu,c\right)$. Therefore, we have set
\begin{equation}
\left( \overline \varphi_\mu^{ac},\varphi_\mu^{ac},\overline \omega_\mu^{ac},\omega_\mu^{ac}\right) =\left( \overline \varphi_i^a,\varphi_i^a,\overline \omega_i^a,\omega_i^a \right)\,.
\end{equation}
The BRST variations of all the fields are given by
\begin{align}\label{BRST1}
sA_{\mu }^{a} &=-\left( D_{\mu }c\right) ^{a}\,, & sc^{a} &=\frac{1}{2}gf^{abc}c^{b}c^{c}\,,   \nonumber \\
s\overline{c}^{a} &=b^{a}\,,&   sb^{a}&=0\,,  \nonumber \\
s\varphi _{i}^{a} &=\omega _{i}^{a}\,,&s\omega _{i}^{a}&=0\,,\nonumber \\
s\overline{\omega}_{i}^{a} &=\overline{\varphi }_{i}^{a}\,,& s \overline{\varphi }_{i}^{a}&=0\,.
\end{align}
However, due to the $\gamma$ dependent term, $S_\gamma$, the Gribov-Zwanziger action breaks  the BRST symmetry softly \cite{Zwanziger:1989mf,Dudal:2008sp}, see eq.~\eqref{breaking}. In order to discuss the renormalizability of $S_\GZ$, we should treat the breaking as a composite operator to be introduced into the action by means of a suitable set of external sources. This procedure can be done in a BRST invariant way, by embedding $S_\GZ$ into a larger action, namely
\begin{eqnarray}\label{brstinvariant}
\Sigma_\GZ &=& S_{\YM} + S_{\gf} + S_0 + S_\s \,,
\end{eqnarray}
whereby
\begin{eqnarray}\label{previous}
S_\s &=& s\int \d^d x \left( -U_\mu^{ai} D_\mu^{ab} \varphi_i^b - V_\mu^{ai} D_{\mu}^{ab} \overline \omega_i^{b} - U_\mu^{ai} V_\mu^{ai}  + T_\mu^{a i} g f_{abc} D^{bd}_\mu c^d \overline \omega^c_i \right)\nonumber\\
&=& \int \d^d x \left( -M_\mu^{ai}  D_\mu^{ab} \varphi_i^b - gf^{abc} U_\mu^{ai}   D^{bd}_\mu c^d  \varphi_i^c   + U_\mu^{ai}  D_\mu^{ab} \omega_i^b - N_\mu^{ai}  D_\mu^{ab} \overline \omega_i^b - V_\mu^{ai}  D_\mu^{ab} \overline \varphi_i^b \right. \nonumber\\
&&\left.+ gf^{abc} V_\mu^{ai} D_\mu^{bd} c^d \overline \omega_i^c - M_\mu^{ai} V_\mu^{ai}+U_\mu^{ai} N_\mu^{ai}  + R_\mu^{ai} g f^{abc} D_\mu^{bd} c^d \overline \omega^c_i  + T_\mu^{ai} g f_{abc} D^{bd}_\mu c^d \overline \varphi^c_i\right) \,. \nonumber\\
\end{eqnarray}
We have introduced 3 new doublets ($U_\mu^{ai}$, $M_\mu^{ai}$), ($V_\mu^{ai}$, $N_\mu^{ai}$) and ($T_\mu^{ai}$, $R_\mu^{ai}$) with the following BRST transformations,
and
\begin{align}
sU_{\mu }^{ai} &= M_{\mu }^{ai}\,, & sM_{\mu }^{ai}&=0\,,  \nonumber \\
sV_{\mu }^{ai} &= N_{\mu }^{ai}\,, & sN_{\mu }^{ai}&=0\,,\nonumber \\
sT_{\mu }^{ai} &= R_{\mu }^{ai}\,, & sR_{\mu }^{ai}&=0\;.
\end{align}
We have therefore restored the broken BRST at the expense of introducing new sources. However, we do not want to alter our original theory \eqref{GZstart}. Therefore, at the end, we have to set the sources equal to the following values:
\begin{eqnarray}\label{physlimit}
&& \left. U_\mu^{ai}\right|_{\phys} = \left. N_\mu^{ai}\right|_{\phys} = \left. T_\mu^{ai}\right|_{\phys} = 0 \,, \nonumber\\
&& \left. M_{\mu \nu }^{ab}\right|_{\phys}= \left.V_{\mu \nu}^{ab}\right|_{\phys}=  -\left.R_{\mu \nu}^{ab}\right|_{\phys} = \gamma^2 \delta ^{ab}\delta _{\mu \nu } \,.
\end{eqnarray}

\subsection{The Ward identities\label{sectwardid}}
Following the procedure of  the algebraic renormalization outlined in \cite{Piguet1995}, we should try to find as many Ward identities as possible. Before doing this, in order to be able to write the Slavnov-Taylor identity, we first have to couple all nonlinear BRST transformations to a new source. Looking at \eqref{BRST1}, we see that only $A_\mu^a$ and $c^a$ transform nonlinearly under the BRST $s$. Therefore, we add the following term to the action $\Sigma_\GZ $,
\begin{eqnarray}\label{ext}
S_{\mathrm{ext}}&=&\int \d^d x\left( -K_{\mu }^{a}\left( D_{\mu }c\right) ^{a}+\frac{1}{2}gL^{a}f^{abc}c^{b}c^{c}\right) \;,
\end{eqnarray}
with $K_{\mu }^{a}$ and $L^a$ two new sources which shall be put to zero at the end,
\begin{eqnarray}\label{physlimit2}
\left. K_{\mu }^{a}\right|_{\phys} =\left. L^{a}\right|_{\phys}  = 0\;.&
\end{eqnarray}
These sources are invariant under the BRST transformation,
\begin{align}
s K_{\mu }^{a} &=0\;, & s L^{a} &= 0\;.
\end{align}
The new action is therefore given by
\begin{eqnarray}\label{enlarged}
\Sigma'_\GZ &=& \Sigma_\GZ + S_{\mathrm{ext}} \;.
\end{eqnarray}
The next step is now to find the Ward identities obeyed by the action $\Sigma'_\GZ$. We have enlisted all the identities below:

\begin{table}
\caption{Quantum numbers of the fields.}
\label{2tabel1}
\begin{center}
\begin{tabular}{|c|c|c|c|c|c|c|c|c|}
\hline
& $A_{\mu }^{a}$ & $c^{a}$ & $\overline{c}^{a}$ & $b^{a}$ & $\varphi_{i}^{a} $ & $\overline{\varphi }_{i}^{a}$ &                $\omega _{i}^{a}$ & $\overline{\omega }_{i}^{a}$   \\
\hline
\textrm{dimension} & $1$ & $0$ &$2$ & $2$ & $1$ & $1$ & $1$ & $1$ \\
$\mathrm{ghost\; number}$ & $0$ & $1$ & $-1$ & $0$ & $0$ & $0$ & $1$ & $-1$ \\
$Q_{f}\textrm{-charge}$ & $0$ & $0$ & $0$ & $0$ & $1$ & $-1$& $1$ & $-1$ \\
\hline
\end{tabular}
\end{center}
\end{table}

\begin{table}
\caption{Quantum numbers of the sources.}
\begin{center}
\label{2tabel2}
\begin{tabular}{|c|c|c|c|c|c|c|c|c|}\hline
        $U_{\mu}^{ai}$&$M_{\mu }^{ai}$&$N_{\mu }^{ai}$&$V_{\mu }^{ai}$& $R_{\mu }^{ai}$  &  $T_{\mu }^{ai}$ &$K_{\mu }^{a}$&$L^{a}$  \\
\hline
         $2$ & $2$ & $2$ &$2$ & 2&2  & $3$ & $4$  \\
         $-1$& $0$ & $1$ & $0$ & 0& -1 & $-1$ & $-2$  \\
         $-1$ & $-1$ & $1$ & $1$ &1&1& $0$ & $0$  \\
\hline
\end{tabular}
\end{center}
\end{table}
\begin{enumerate}
\item The Slavnov-Taylor identity is given by
\begin{equation}\label{slavnov}
\mathcal{S}(\Sigma'_\GZ )=0\;,
\end{equation}
with
\begin{multline*}
\mathcal{S}(\Sigma'_\GZ ) =\int \d^d x\left( \frac{\delta \Sigma'_\GZ
}{\delta K_{\mu }^{a}}\frac{\delta \Sigma'_\GZ }{\delta A_{\mu
}^{a}}+\frac{\delta \Sigma'_\GZ }{\delta L^{a}}\frac{\delta \Sigma'_\GZ
}{\delta c^{a}} \right. \nonumber\\
\left. +b^{a}\frac{\delta \Sigma'_\GZ }{\delta \overline{c}^{a}}+\overline{\varphi }_{i}^{a}\frac{\delta \Sigma'_\GZ }{\delta \overline{\omega }_{i}^{a}}+\omega _{i}^{a}\frac{\delta \Sigma'_\GZ }{\delta \varphi _{i}^{a}} +M_{\mu }^{ai}\frac{\delta \Sigma'_\GZ}{\delta U_{\mu}^{ai}}+N_{\mu }^{ai}\frac{\delta \Sigma'_\GZ }{\delta V_{\mu }^{ai}} + R_{\mu }^{ai}\frac{\delta \Sigma'_\GZ }{\delta T_{\mu }^{ai}}\right) \;.
\end{multline*}

\item The $U(f)$ invariance is given by
\begin{equation}
U_{ij} \Sigma'_\GZ =0\;,\label{ward1}
\end{equation}
\begin{multline}
U_{ij}=\int \d^dx\Bigl( \varphi_{i}^{a}\frac{\delta }{\delta \varphi _{j}^{a}}-\overline{\varphi}_{j}^{a}\frac{\delta }{\delta \overline{\varphi}_{i}^{a}}+\omega _{i}^{a}\frac{\delta }{\delta \omega _{j}^{a}}-\overline{\omega }_{j}^{a}\frac{\delta }{\delta \overline{\omega }_{i}^{a}} \\
-  M^{aj}_{\mu} \frac{\delta}{\delta M^{ai}_{\mu}} -U^{aj}_{\mu}\frac{\delta}{\delta U^{ai}_{\mu}} + N^{ai}_{\mu}\frac{\delta}{\delta N^{aj}_{\mu}}
  +V^{ai}_{\mu}\frac{\delta}{\delta V^{aj}_{\mu}}    +  R^{aj}_{\mu}\frac{\delta}{\delta R^{ai}_{\mu}} + T^{aj}_{\mu}\frac{\delta}{\delta T^{ai}_{\mu}} \Bigr)  \;. \nonumber
\end{multline}
By means of the diagonal operator $Q_{f}=U_{ii}$, the $i$-valued fields and sources can be assigned an additional charge.
One can find all quantum numbers in Table \ref{2tabel1} and Table \ref{2tabel2}.

\item The Landau gauge condition reads
\begin{eqnarray}\label{gaugeward}
\frac{\delta \Sigma'_\GZ }{\delta b^{a}}&=&\partial_\mu A_\mu^{a}\;.
\end{eqnarray}

\item The antighost equation yields
\begin{eqnarray}
\frac{\delta \Sigma'_\GZ }{\delta \overline{c}^{a}}+\partial _{\mu}\frac{\delta \Sigma'_\GZ }{\delta K_{\mu }^{a}}&=&0\;.
\end{eqnarray}

\item The linearly broken local constraints yield
\begin{eqnarray}
\frac{\delta \Sigma'_\GZ }{\delta \overline{\varphi }^{a}_i}+\partial _{\mu }\frac{\delta \Sigma'_\GZ }{\delta M_{\mu }^{ai}} + g f_{dba}    T^{d i}_\mu \frac{\delta \Sigma'_\GZ }{\delta K_{\mu }^{b i}} &=&gf^{abc}A_{\mu }^{b}V_{\mu}^{ci} \;, \nonumber\\
\frac{\delta \Sigma'_\GZ }{\delta \omega ^{a}_i}+\partial _{\mu}\frac{\delta \Sigma'_\GZ }{\delta N_{\mu}^{ai}}-gf^{abc}\overline{\omega }^{b}_i\frac{\delta \Sigma'_\GZ }{\delta b^{c}}&=&gf^{abc}A_{\mu }^{b}U_{\mu }^{ci}\;. \;.
\end{eqnarray}

\item The exact $\mathcal{R}_{ij}$ symmetry reads
\begin{equation}
\mathcal{R}_{ij}\Sigma'_\GZ =0\;,
\end{equation}
with
\begin{multline}\label{rij}
\mathcal{R}_{ij} = \int \d^dx\left( \varphi _{i}^{a}\frac{\delta}{\delta\omega _{j}^{a}}-\overline{\omega }_{j}^{a}\frac{\delta }{\delta \overline{\varphi }_{i}^{a}}+V_{\mu }^{ai}\frac{\delta }{\delta N_{\mu}^{aj}}-U_{\mu }^{aj}\frac{\delta }{\delta M_{\mu }^{ai}} + T^{a i}_\mu \frac{\delta }{\delta R_{\mu }^{aj}}  \right) \;.
\end{multline}

\item The integrated Ward identity is given by
\begin{equation}
\int \d^d x \left( c^a \frac{ \delta \Sigma'_\GZ }{ \delta \omega^{ a}_i} + \overline \omega^{a}_i \frac{ \delta \Sigma'_\GZ }{ \delta  \overline c^a } + U^{a i}_\mu \frac{ \delta \Sigma'_\GZ }{ \delta  K^a_\mu }  \right) = 0\;.
\end{equation}
\end{enumerate}
Here we should add that due to the presence of the sources $T_{\mu }^{ai}$ and $R_{\mu }^{ai}$, the powerful ghost  Ward identity \cite{Piguet1995} is broken, and we are unable to restore this identity. For the standard Yang-Mills theory, this identity has the following form
\begin{equation}\label{GW}
\mathcal{G}^{a}( S_\YM + S_\gf)=\Delta _{\mathrm{cl}}^{a}\,,
\end{equation}
with
\begin{eqnarray}
\mathcal{G}^{a} &=&\int \d^dx\left( \frac{\delta }{\delta c^{a}}+gf^{abc} \overline{c}^{b}\frac{\delta }{\delta b^{c}} \right) \,,
\end{eqnarray}
and
\begin{equation}
\Delta _{\mathrm{cl}}^{a}=g\int \d^{4}xf^{abc}\left( K_{\mu}^{b}A_{\mu }^{c}-L^{b}c^{c}\right) \;,
\end{equation}
i.e.~a linear breaking. However, it shall turn out that this is not a problem for the renormalization procedure being undertaken, see later.

\subsection{The counterterm\label{sectcounterterm}}
The next step in the algebraic renormalization is to translate all these symmetries, which are not anomalous, into constraints on the counterterm $\Sigma_\GZ^c$, which is an integrated polynomial in the fields and sources of dimension four and with ghost number zero. The classical action $\Sigma_\GZ'$ changes under quantum corrections according to
\begin{eqnarray}
    \Sigma_\GZ' \rightarrow \Sigma_\GZ' + h \Sigma_\GZ^c\,,
\end{eqnarray}
whereby $h$ is the perturbation parameter. Demanding that the perturbed action $(\Sigma_\GZ' + h \Sigma_\GZ^c)$ fulfills the same set of Ward identities obeyed by $\Sigma_\GZ'$, it follows that the counterterm $\Sigma_\GZ^c$ is constrained by the following identities:
\begin{enumerate}
\item The linearized Slavnov-Taylor identity yields
\begin{equation}
\mathcal{B}\Sigma_\GZ^{c}=0\;,
\end{equation}
with $\mathcal{B}$ the nilpotent linearized Slavnov-Taylor operator,
\begin{multline}
\mathcal{B}=\int \d^{4}x\Bigl( \frac{\delta \Sigma_\GZ'}{\delta K_{\mu }^{a}}\frac{\delta }{\delta A_{\mu }^{a}}+\frac{\delta \Sigma_\GZ' }{\delta A_{\mu }^{a}}\frac{\delta }{\delta K_{\mu }^{a}}+\frac{\delta \Sigma_\GZ' }{\delta L^{a}}\frac{\delta }{\delta c^{a}}+\frac{\delta\Sigma_\GZ' }{\delta c^{a}}\frac{\delta }{\delta L^{a}}+b^{a}\frac{\delta }{\delta \overline{c}^{a}}\\
+\overline{\varphi}_{i}^{a}\frac{\delta }{\delta \overline{\omega }_{i}^{a}}+\omega_{i}^{a}\frac{\delta }{\delta \varphi_{i}^{a}}
+M_{\mu }^{ai}\frac{\delta }{\delta U_{\mu }^{ai}} + N_{\mu }^{ai}\frac{\delta }{\delta V_{\mu }^{ai}} + R_{\mu }^{ai}\frac{\delta }{\delta T_{\mu }^{ai}}  \Bigr) \,,
\end{multline}
and
\begin{equation}
\mathcal{B}^2=0\;.
\end{equation}

\item The $U(f)$ invariance gives
\begin{eqnarray}
U_{ij} \Sigma_\GZ^{c} &=&0 \;.
\end{eqnarray}

\item The Landau gauge condition
\begin{eqnarray}
\frac{\delta \Sigma_\GZ^{c}}{\delta b^{a}}&=&0\,.
\end{eqnarray}

\item The antighost equation
\begin{eqnarray}
\frac{\delta \Sigma_\GZ^{c}}{\delta \overline c^{a}}+\p_\mu\frac{\delta \Sigma_\GZ^{c}}{\delta K_{\mu}^a} &=&0 \,.
\end{eqnarray}

\item The linearly broken local constraints yield
\begin{eqnarray}
\left( \frac{\delta  }{\delta \overline{\varphi }^{a}_i}+\partial _{\mu }\frac{\delta}{\delta M_{\mu }^{ai}} +\partial _{\mu }\frac{\delta }{\delta M_{\mu }^{ai}} + g f_{abc}    T^{b i}_\mu \frac{\delta }{\delta K_{\mu }^{c i}}\right) \Sigma_\GZ^{c} &=&0 \;,\nonumber\\
\left( \frac{\delta }{\delta \omega ^{a}_i}+\partial _{\mu}\frac{\delta }{\delta N_{\mu}^{ai}}-gf^{abc}\overline{\omega }^{b}_i \frac{\delta }{\delta b^{c}} \right) \Sigma_\GZ^{c}&=&0 \;.
\end{eqnarray}

\item The exact $\mathcal{R}_{ij}$ symmetry imposes
\begin{equation}
\mathcal{R}_{ij}\Sigma_\GZ^{c}=0\;,
\end{equation}
with  $\mathcal{R}_{ij}$ given in \eqref{rij}.

\item Finally, the integrated Ward identity becomes
\begin{equation}
\int \d^d x \left( c^a \frac{ \delta \Sigma_\GZ^{c}}{ \delta \omega^{ a}_i} + \overline \omega^{a}_i \frac{ \delta \Sigma_\GZ^{c}}{ \delta  \overline c^a } + U^{a i}_\mu \frac{ \delta \Sigma_\GZ^{c}}{ \delta  K^a_\mu }  \right) = 0 \;.
\end{equation}
\end{enumerate}
 The most general counterterm $\Sigma_\GZ^{c}$ of $d=4$, which obeys the linearized Slavnov-Taylor identity, has ghost number zero, and vanishing $Q_f$ number,  can be written as
\begin{eqnarray}
\Sigma^c_\GZ &=& a_0 S_{\YM} + \mathcal{B} \int \d^d \!x\,   \biggl\{  a_{1} K_{\mu}^{a} A_{\mu}^{a} + a_2 \partial _{\mu} \overline{c}^{a} A_{\mu}^{a}+a_3 \,L^{a}c^{a}
+a_4 U_{\mu}^{ai}\,\partial _{\mu }\varphi _{i}^{a} +a_5 \,V_{\mu}^{ai}\,\partial _{\mu }\overline{\omega }_{i}^{a}\nonumber\\
&& + a_6 \overline{\omega}_{i}^{a} \partial^{2} \varphi_{i}^{a} + a_7 U_{\mu}^{ai}V_{\mu}^{ai} + a_8 gf^{abc}U_{\mu}^{ai}\varphi_{i}^{b}A_{\mu}^{c}+a_9 gf^{abc}V_{\mu}^{ai}\overline{\omega }_{i}^{b}A_{\mu }^{c}\nonumber \\
&& +a_{10}gf^{abc}\overline{\omega }_{i}^{a}A_{\mu }^{c}\,\partial _{\mu }\varphi _{i}^{b}+a_{11}gf^{abc}\overline{\omega }_{i}^{a}(\partial _{\mu }A_{\mu}^{c})\varphi _{i}^{b} +b_1 R_{\mu}^{ai} U_{\mu}^{ai} +b_2 T_{\mu }^{ai} M_{\mu }^{ai} \nonumber\\
&&+ b_3 g f_{abc} R_{\mu }^{ai} \overline{\omega }_{i}^{b} A_{\mu}^{c} + b_4 g f_{abc} T_{\mu}^{ai} \overline{\varphi }_{i}^{b} A_{\mu}^{c} + b_5 R_{\mu}^{ai} \p_\mu \overline{\omega }_{i}^{a}  + b_6 T_{\mu}^{ai} \p_\mu \overline{\varphi }_{i}^{a} \biggr\} \;,
\end{eqnarray}
with $a_0, \ldots, a_{11}, b_1, \ldots, b_6$ arbitrary parameters. Now we can impose the constraints on the counterterm. Firstly, although the ghost Ward identity \eqref{GW} is broken, we know that this is not so in the standard Yang-Mills case. Therefore, we can already set $a_3=0$ as this term is not allowed in the counterterm of the standard Yang-Mills action, which is a special case of the action we are studying\footnote{In particular, since we will always assume the use of a mass independent renormalization scheme, we may compute $a_3$ with all external mass scales (= sources) equal to zero. Said otherwise, $a_3$ is completely determined by the dynamics of the original Yang-Mills action, in which case it is known to vanish to all orders.}. Secondly, due to the Landau gauge condition (3.) and the antighost equation (4.) we find,
\begin{eqnarray}
a_1 &=& a_2\;.
\end{eqnarray}
Next, the linearly broken constraints (5.) give the following relations
\begin{align}
 a_1 &= -a_8  = - a_{9} = a_{10} = a_{11} = -b_3 = b_4\;, \nonumber\\
  a_4 &= a_5 = -a_6 = a_7\;, \quad b_1 =b_2 = b_5 = b_6 = 0 \;.
\end{align}
The $R_{ij}$ symmetry (6.) does not give any new information, while the integrated Ward identity (7.) relates the two previous strings of parameters:
\begin{multline}
 a_1 = -a_8  = - a_{9} = a_{10} = a_{11} = -b_3 = b_4  \equiv     a_3 = a_4 = -a_5 = a_6 \;.
\end{multline}
Taking all this information together, we obtain the following counterterm
\begin{multline}\label{final}
\Sigma^c= a_{0}S_{YM}  + a_{1}\int \d^dx\Biggl(  A_{\mu}^{a}\frac{ \delta S_{YM}}{\delta A_{\mu }^{a}}  + \p_\mu \overline{c}^a \p_\mu c^a  + K_{\mu }^{a}\partial _{\mu }c^{a}  + M_\mu^{a i} \p_\mu \varphi^{a}_i -  U_\mu^{a i} \p_\mu \omega^{a}_i \\
+ N_\mu^{a i} \p_\mu \overline{\omega}_i^{a} +  V_\mu^{a i}\p_\mu \overline{\varphi}^{a}_i   +  \p_\mu \overline{\varphi}^{a}_i \p_\mu \varphi^{a}_i +  \p_\mu \omega^{a}_i \p_\mu \overline{\omega}^{a}_i + V_\mu^{a i} M_\mu^{a i} - U_\mu^{a i}N_\mu^{a i}  - g f_{abc} U_\mu^{ai} \varphi^{b}_i \p_\mu c^c \\
- g f_{abc} V_\mu^{ai} \overline{\omega}^{b}_i \p_\mu c^c - g f_{abc} \p_{\mu} \overline{\omega}^a_i \varphi^{b}_i  \p_\mu c^c   - g f_{abc} R_\mu^{ai} \p_\mu c^b \overline \omega_i^c + g f_{abc} T_\mu^{ai} \p_\mu c^b \overline \varphi_i^c \Biggr) \;.
\end{multline}

\subsection{The renormalization factors}
As a final step, we have to show that the counterterm \eqref{final} can be reabsorbed by means of a multiplicative renormalization of the fields and sources.
If we try to absorb the counterterm into the original action, we easily find,
\begin{eqnarray}\label{Z1}
Z_{g} &=&1-h \frac{a_0}{2}\,,  \nonumber \\
Z_{A}^{1/2} &=&1+h \left( \frac{a_0}{2}+a_{1}\right) \,,
\end{eqnarray}
and
\begin{eqnarray}\label{Z2}
Z_{\overline{c}}^{1/2} &=& Z_{c}^{1/2} = Z_A^{-1/4} Z_g^{-1/2} = 1-h \frac{a_{1}}{2}\,, \nonumber \\
Z_{b}&=&Z_{A}^{-1}\,, \nonumber\\
Z_{K }&=&Z_{c}^{1/2}\,,  \nonumber\\
Z_{L} &=&Z_{A}^{1/2}\,.
\end{eqnarray}
The results \eqref{Z1} are already known from the renormalization of the original Yang-Mills action in the Landau gauge \cite{Piguet1995}. Further, we also obtain
\begin{eqnarray}\label{Z3}
Z_{\varphi}^{1/2} &=& Z_{\overline \varphi}^{1/2} = Z_g^{-1/2} Z_A^{-1/4} = 1 - h \frac{a_1}{2}\,, \nonumber\\
Z_\omega^{1/2} &=& Z_A^{-1/2} \,,\nonumber\\
Z_{\overline \omega}^{1/2} &=& Z_g^{-1} \,,\nonumber\\
Z_M &=& 1- \frac{a_1}{2} = Z_g^{-1/2} Z_A^{-1/4}\,, \nonumber\\
Z_N &=& Z_A^{-1/2} \,, \nonumber\\
Z_U &=& 1 + h \frac{a_0}{2} = Z_g^{-1} \,, \nonumber\\
Z_V &=& 1- h \frac{a_1}{2} = Z_g^{-1/2}Z_A^{-1/4} \,,  \nonumber\\
Z_T &=& 1+h \frac{a_0}{2} = Z_g^{-1}  \,,  \nonumber\\
Z_R &=& 1- h \frac{a_1}{2} = Z_g^{-1/2}Z_A^{-1/4}\;.
\end{eqnarray}
This concludes the proof of the renormalizability of the action \eqref{GZstart} which is the physical limit of $\Sigma_\GZ' $. Notice that in the physical limit \eqref{physlimit}, we have that
\begin{eqnarray}\label{Zgamma}
Z_{\gamma^2} &=& Z_g^{-1/2} Z_A^{-1/4}\;.
\end{eqnarray}

\section{Inclusion of the operator $A^2$ in the Gribov-Zwanziger action\label{appendixB}}
For the benefit of the reader, let us also repeat the renormalization of the operator $A^2$ in the Gribov-Zwanziger action, which was first tackled in \cite{Dudal:2005na}. In this paper, it was shown that the presence of the condensate $\braket{A^2}$ does not spoil the renormalizability of the GZ action. The GZ action with inclusion of the local composite operator $A_\mu^a A_\mu^a$ is given by
\begin{eqnarray}
S_\AGZ &=& S_\GZ + S_{A^2}\;,
\end{eqnarray}
whereby
\begin{eqnarray}
S_{A^2} &=& \int \d^d x \left( \frac{\tau }{2}A_{\mu }^{a}A_{\mu }^{a} - \frac{\zeta }{2}\tau ^{2}\right)\;,
\end{eqnarray}
with $\tau$ a new source invariant under the BRST transformation $s$ and $\zeta$ a new parameter. The renormalization can be done very easily with the help of the previous section.

\subsection{The starting action and the BRST}
Again, we shall make $S_\AGZ$ BRST invariant. We define
\begin{eqnarray}\label{AGZ}
\Sigma_\AGZ &=& \Sigma'_\GZ + \Sigma_{A^2}
\end{eqnarray}
whereby $\Sigma'_\GZ$ is given in expression \eqref{enlarged} and
\begin{eqnarray}\label{sigmaAkwadraat}
\Sigma_{A^2} &=& \int \d^{4}x\; s\left( \frac{\eta }{2}A_{\mu }^{a}A_{\mu}^{a}-\frac{\zeta }{2}\tau ^{2}\right)~=~ \int \d^4 x \left[\frac{1}{2} \tau A_{\mu }^{a}A_{\mu }^{a}+\eta A_{\mu }^{a}\partial _{\mu }c^{a}-\frac{1}{2}\zeta \tau ^{2} \right]
\end{eqnarray}
with $\eta$ a new source and $s \eta = \tau$, so that $(\eta, \tau)$ forms a doublet.  At the end, we replace all the sources with their physical values, see expression \eqref{physlimit} and \eqref{physlimit2}, and in addition
\begin{eqnarray}
 &\left. \eta \right|_{\phys} = 0\;,&
\end{eqnarray}
so one recovers $S_\AGZ$ again.

\subsection{The Ward identities}
It is now easily checked that the Ward identities 1-7 of section \ref{sectwardid} remain preserved. Obviously, the Slavnov-Taylor identity receives an extra term,
\begin{equation}
\mathcal{S}(\Sigma_\AGZ )=0\;,
\end{equation}
whereby
\begin{multline*}
\mathcal{S}(\Sigma_\AGZ ) =\int \d^d x\left( \frac{\delta \Sigma_\AGZ }{\delta K_{\mu }^{a}}\frac{\delta \Sigma_\AGZ}{\delta A_{\mu}^{a}}+\frac{\delta \Sigma_\AGZ  }{\delta L^{a}}\frac{\delta \Sigma_\AGZ}{\delta c^{a}} \right. \nonumber\\
\left. +b^{a}\frac{\delta \Sigma_\AGZ }{\delta \overline{c}^{a}}+\overline{\varphi }_{i}^{a}\frac{\delta \Sigma_\AGZ }{\delta \overline{\omega }_{i}^{a}}+\omega _{i}^{a}\frac{\delta \Sigma_\AGZ }{\delta \varphi _{i}^{a}} +M_{\mu }^{ai}\frac{\delta \Sigma_\AGZ }{\delta U_{\mu}^{ai}}+N_{\mu }^{ai}\frac{\delta \Sigma_\AGZ}{\delta V_{\mu }^{ai}} + R_{\mu }^{ai}\frac{\delta \Sigma_\AGZ }{\delta T_{\mu }^{ai}} + \tau  \frac{\delta \Sigma_\AGZ }{\delta \eta} \right) \;.
\end{multline*}

\subsection{The counterterm}
As all the Ward identities remain the same, it is easy to check that the counterterm is given by
\begin{eqnarray}\label{countertermAGZ}
\Sigma^c_\AGZ &=& \Sigma^c_\GZ + \int \d^{4}x\left( \frac{a_{2}}{2}\tau A_{\mu }^{a}A_{\mu }^{a}+ \frac{a_{3}}{2}\zeta \tau ^{2}+\left( a_{2}-a_{1}\right) \eta A_{\mu
}^{a}\partial _{\mu }c^{a}\right) \;.
\end{eqnarray}
whereby $\Sigma^c_\GZ$ is the counterterm \eqref{final}. This counterterm can be absorbed in the original action, $\Sigma_\AGZ$ leading to the same renormalization factors as in equations \eqref{Z1}-\eqref{Z3}.  \\
\\
In addition $Z_{\tau}$ is related to $Z_{g}$ and $Z_{A}^{1/2}$ \cite{Dudal:2005na}:
\begin{eqnarray}
    Z_\tau=Z_gZ_A^{-1/2}\;,
\end{eqnarray}
and $Z_\zeta$ and $Z_\eta$ are given by
\begin{eqnarray}
Z_{\zeta }&=&1+ h (-a_{3}-2a_{2}+4a_{1}-2a_{0}) \;,\nonumber\\
Z_\eta &=& 1 + h (\frac{a_0}{2} - \frac{3}{2}a_1 + a_2 )\;.
\end{eqnarray}

\section{Renormalization of the further refined action\label{appendixC}}

\subsection{The starting action}
Let us repeat the starting action \eqref{CGZ},
\begin{eqnarray}\label{actionCGZ}
\Sigma_\CGZ &=& \Sigma_\GZ'  + \Sigma_{A^2} + S_{\varphi \overline \varphi}  + S_{\overline \omega \omega} + S_{\overline{\varphi} \overline \varphi, \overline \omega \overline \varphi }  + S_{\varphi \varphi, \omega \varphi } + S_\vac \;,
\end{eqnarray}
whereby $\Sigma_\GZ'$ is given by equation \eqref{enlarged}, $\Sigma_{A^2}$ by \eqref{sigmaAkwadraat} and
\begin{eqnarray}
S_{\varphi \overline \varphi } & =&  \int \d^4 x s(  P \overline \varphi^a_i \varphi^a_i ) ~=~  \int \d^4 x \left[  Q  \overline \varphi^a_i \varphi^a_i- P  \overline \varphi^a_i \omega^a_i\right]  \;, \nonumber\\
S_{\overline \omega \omega } & =&  \int \d^4 x s(  V \overline \omega^a_i \omega^a_i) ~=~  \int \d^4 x \left[ W\overline \omega^a_i \omega^a_i - V \overline \varphi^a_i \omega^a_i \right]  \;, \nonumber\\
S_{\overline{\varphi} \overline \varphi, \overline \omega \overline \varphi } & =&  \frac{1}{2} \int \d^4 x s(  \overline G^{ij} \overline \omega^a_i \overline \varphi^a_j) ~=~  \int \d^4 x \left[  \overline H^{ij} \overline \omega^a_i \overline \varphi^a_j + \frac{1}{2} \overline G^{ij} \overline \varphi^a_i \overline \varphi^a_j \right]  \;, \nonumber\\
S_{\varphi \varphi, \omega \varphi } &=& \frac{1}{2} \int \d^4 x s(  H^{ij} \varphi^a_i  \varphi^a_j) ~=~   \int \d^4 x  \left[ \frac{1}{2} G^{ij} \varphi^a_i \varphi^a_j - H^{ij} \omega^a_i \varphi^a_j \right] \;, \nonumber\\
S_\vac &=&  \int \d^4 x \left[ \kappa (G^{ij} \overline G^{ij} - 2 H^{ij} \overline H^{ij}) +  \lambda (G^{ii} \overline G^{jj} - 2 H^{ii} \overline H^{jj}) \right] \nonumber\\
 && - \int \d^4 x \left[ \alpha   (Q Q +  Q W) + \beta ( QW + W W)  + \chi Q \tau + \delta W \tau \right] \;.
\end{eqnarray}

\subsection{The Ward identities}
With the help of appendix \ref{appendixA}, we can easily summarize all Ward identities obeyed by the action $\Sigma_\CGZ$
\begin{enumerate}
\item  The Slavnov-Taylor identity reads
\begin{equation}\label{qqf1}
\mathcal{S}(\Sigma_\CGZ )=0\;,
\end{equation}
with
\begin{eqnarray*}
\mathcal{S}(\Sigma_\CGZ ) &=&\int \d^{4}x\left( \frac{\delta \Sigma_\CGZ}{\delta K_{\mu }^{a}}\frac{\delta \Sigma_\CGZ }{\delta A_{\mu}^{a}}+\frac{\delta \Sigma_\CGZ }{\delta L^{a}}\frac{\delta \Sigma_\CGZ}{\delta c^{a}} + b^{a}\frac{\delta \Sigma_\CGZ}{\delta \overline{c}^{a}}+\overline{\varphi }_{i}^{a}\frac{\delta \Sigma_\CGZ }{\delta \overline{\omega }_{i}^{a}} \right. \\
&&+\left.\omega _{i}^{a}\frac{\delta \Sigma_\CGZ }{\delta \varphi _{i}^{a}} +M_{\mu }^{ai}\frac{\delta \Sigma_\CGZ}{\delta U_{\mu}^{ai}} +N_{\mu }^{ai}\frac{\delta \Sigma_\CGZ }{\delta V_{\mu }^{ai}} + R_{\mu }^{ai}\frac{\delta \Sigma_\CGZ }{\delta T_{\mu }^{ai}}  +  Q \frac{\delta \Sigma_\CGZ }{\delta P} \right.\nonumber\\
&&\left.+  W \frac{\delta \Sigma_\CGZ }{\delta V} + \tau \frac{\delta \Sigma_\CGZ }{\delta \eta} + 2  \overline H^{ij} \frac{\delta \Sigma_\CGZ }{\delta \overline G^{ij}} +  G^{ij} \frac{\delta \Sigma_\CGZ }{\delta H^{ij}} \right) \;.
\end{eqnarray*}

\item For the $U(f)$ invariance we now have
\begin{eqnarray}
U_{ij} \Sigma_\CGZ &=&0\;,
\end{eqnarray}
whereby
\begin{multline*}
U_{ij}=\int \d^{4}x\left( \varphi_{i}^{a}\frac{\delta }{\delta \varphi _{j}^{a}}-\overline{\varphi
}_{j}^{a}\frac{\delta }{\delta \overline{\varphi}_{i}^{a}}+\omega _{i}^{a}\frac{\delta }{\delta \omega _{j}^{a}}-\overline{\omega }_{j}^{a}\frac{\delta }{\delta \overline{\omega }_{i}^{a}}
-  M^{aj}_{\mu} \frac{\delta}{\delta M^{ai}_{\mu}} -U^{aj}_{\mu}\frac{\delta}{\delta U^{ai}_{\mu}} +N^{ai}_{\mu}\frac{\delta}{\delta N^{aj}_{\mu}}  \right.\\
\left. +V^{ai}_{\mu}\frac{\delta}{\delta V^{aj}_{\mu}}+ R^{aj}_{\mu}\frac{\delta}{\delta R^{ai}_{\mu}} + T^{aj}_{\mu}\frac{\delta}{\delta T^{ai}_{\mu}} +2 \overline G^{ki} \frac{\delta}{\delta \overline G^{kj}}  -2 G^{kj} \frac{\delta}{\delta G^{ki}} + 2 \overline H ^{ki} \frac{\delta}{\delta \overline H^{kj}} -2  H^{kj} \frac{\delta}{\delta H^{ki}  } \right) \;.
\end{multline*}
By means of the diagonal operator $Q_{f}=U_{ii}$, the single $i$-valued fields and sources still turn out to possess an additional quantum number.

\item  The Landau gauge condition and the antighost equation are given by
\begin{eqnarray}
\frac{\delta \Sigma_\CGZ }{\delta b^{a}}&=&\partial_\mu A_\mu^{a}\;,\\
\frac{\delta \Sigma_\CGZ }{\delta \overline{c}^{a}}+\partial _{\mu
}\frac{\delta \Sigma_\CGZ }{\delta K_{\mu }^{a}}&=&0\;.
\end{eqnarray}

\item  The linearly broken local constraints yield
\begin{eqnarray}
\frac{\delta \Sigma_\CGZ }{\delta \overline{\varphi }^{a}_i}+\partial _{\mu }\frac{\delta \Sigma_\CGZ }{\delta M_{\mu }^{ai}} + g f_{dba}    T^{d i}_\mu \frac{\delta \Sigma_\CGZ }{\delta K_{\mu }^{b i}} &=&gf^{abc}A_{\mu }^{b}V_{\mu}^{ci} + \ldots \;, \nonumber\\
\frac{\delta \Sigma_\CGZ }{\delta \omega ^{a}_i}+\partial _{\mu}\frac{\delta \Sigma_\CGZ }{\delta N_{\mu}^{ai}}-gf^{abc}\overline{\omega }^{b}_i\frac{\delta \Sigma_\CGZ }{\delta b^{c}}&=&gf^{abc}A_{\mu }^{b}U_{\mu }^{ci} + \ldots\;.
\end{eqnarray}
whereby the $\ldots$ are extra linear breaking terms irrelevant for our purposes.

\item  The exact $\mathcal{R}_{ij}$ symmetry is \textbf{broken} beyond simple repair.

\item The integrated Ward Identity is \textbf{broken} also beyond simple repair.

\item There is however a new identity:
\begin{eqnarray}\label{qqf8}
\frac{\delta \Sigma_\CGZ }{\delta P} &=& \frac{\delta \Sigma_\CGZ }{\delta V}\;.
\end{eqnarray}

\end{enumerate}
\begin{table}[H]
        \begin{tabular}{|c|c|c|c|c|c|c|c|c||c|c|c|c|}
        \hline
        & $A_{\mu }^{a}$ & $c^{a}$ & $\overline{c}^{a}$ & $b^{a}$ & $\varphi_{i}^{a} $ & $\overline{\varphi }_{i}^{a}$ & $\omega _{i}^{a}$ & $\overline{\omega }_{i}^{a}$ &$U_{\mu}^{ai}$&$M_{\mu }^{ai}$&$N_{\mu }^{ai}$&$V_{\mu }^{ai}$  \\
        \hline
        \hline
        \textrm{dimension} & $1$ & $0$ &$2$ & $2$ & $1$ & $1$ & $1$ & $1$ &$2$ & $2$ & $2$ &$2$  \\
        \hline
        $\mathrm{ghost\; number}$ & $0$ & $1$ & $-1$ & $0$ & $0$ & $0$ & $1$ & $-1$ & $-1$& $0$ & $1$ & $0$ \\
        \hline
        $Q_{f}\textrm{-charge}$ & $0$ & $0$ & $0$ & $0$ & $1$ & $-1$& $1$ & $-1$&  $-1$& $-1$&$1$  & $1$ \\
        \hline
        \end{tabular}
        \end{table}
        \begin{table}[H]
        \begin{tabular}{|c|c|c|c|c|c|c|c|c|c|c|c|c|c|c|}
        \hline
        &$R_{\mu }^{ai}$&$T_{\mu }^{ai}$&$K_{\mu }^{a}$&$L^{a}$ & $Q$  & $P$ & $W$ & $V$  & $\tau$ & $\eta$ & $G^{ij}$ &$\overline G^{ij}$ &$H^{ij}$&$\overline H^{ij}$ \\
        \hline
        \hline
         \textrm{dimension} &$2$ & $2$& $3$ & $4$ & $2$ & 2  &2&2 & 2&2 &2&2&2&2\\
        \hline
        $\mathrm{ghost\; number}$ &$0$ & $-1$ & $-1$ & $-2$ & 0 &$-1$ &0&$-1 $&0&$-1$&0&0&$-1$&1\\
        \hline
        $Q_{f}\textrm{-charge}$ &1 & 1 & $0$  & $0$ &0 &0 &0&0 &0&0&$-2$&2&$-2$&2\\
        \hline
        \end{tabular}
        \caption{Quantum numbers of the fields and sources.}
        \end{table}

\subsection{The counterterm}
These identities \eqref{qqf1}-\eqref{qqf8} can be translated into constraints on the counterterm according to the quantum action principe (QAP), see \cite{Piguet1995}. Unfortunately, many identities are broken due to the introduction of these $d=2$ operators. However, we are using mass independent renormalization schemes and therefore, the new massive sources ($P$, $Q$, $V$, $W$, $G^{ij}$, $\overline G^{ij}$, $H^{ij}$, $\overline H^{ij}$) cannot influence the counterterm of the original GZ action \eqref{final} since they are coupled to $d=2$ operators. Said otherwise, there are no new vertices capable of destroying the UV-structure of the original GZ theory \eqref{final}. We only need to check whether these operators themselves are renormalizable. Thus, the counterterm is given by
\begin{equation}
\Sigma^c_\CGZ = \Sigma^c_\GZ + \Sigma^c_A + \Sigma^c_{P - H}\;,
\end{equation}
with $\Sigma^c_\GZ$ given by equation \eqref{final}, and $\Sigma^c_A$ given by
\begin{equation}
\Sigma^c_A = \int \d^{4}x\left( \frac{a_{2}}{2}\tau A_{\mu }^{a}A_{\mu }^{a}+ \frac{a_{3}}{2}\zeta \tau ^{2}+\left( a_{2}-a_{1}\right) \eta A_{\mu}^{a}\partial _{\mu }c^{a}\right)\;,
\end{equation}
as already determined in \eqref{countertermAGZ}. $\Sigma^c_{P \ldots H}$ is dependent of all the sources  ($P$, $Q$, $V$, $W$, $G^{ij}$, $\overline G^{ij}$, $H^{ij}$, $\overline H^{ij}$), is of dimension 4, ghost number $-1$ and $Q_f = 0$ and obeys the remaining Ward identities. Due to the linearly broken constraints we find
\begin{align}
\frac{ \p \Sigma^c_{P - H}}{ \p \varphi} &= 0\;, & \frac{ \p \Sigma^c_{P - H}}{ \p \overline \varphi} &= 0 \;, & \frac{ \p \Sigma^c_{P - H}}{ \p \omega} &= 0 \;, &\frac{ \p \Sigma^c_{P - H}}{ \p \overline \omega } &= 0\;.
\end{align}
Therefore,
\begin{multline}
\Sigma^c_{P -H}=  \mathcal B_\Sigma  \int d^4x\; \bigl( b_1 P A_\mu^a A_\mu^a + b_2 V A_\mu^a A_\mu^a + b_3 Q P + b_4 Q V + b_5 W P + b_6 W V + b_7 P \tau + b_8 V \tau \\+ b_9 Q \eta + b_{10} W \eta + c_1 H^{ij} \overline G^{ij} +  c_2 H^{ii}  \overline  G^{jj}  \bigr)\;,
\end{multline}
whereby $b_1$, $\ldots$, $c_2$ are arbitrary constants. By invoking the new identity
\begin{eqnarray}
\frac{\delta\Sigma^c_{P -H} }{\delta P} &=& \frac{\delta \Sigma^c_{P -H} }{\delta V}\;,
\end{eqnarray}
we can write
\begin{multline}
\Sigma^c_{P -H}=   \int d^4x\; b_1 [ (Q+W) A_\mu^a A_\mu^a +2 (P+V) \p_\mu c^a A_\mu^a] + b_3 Q Q + b_4 Q W + b_6 W W  + b_7 Q \tau + b_8 W \tau \\
+  c_1 (G^{ij} \overline G^{ij} - 2 H^{ij} \overline H^{ij}) + c_{2}   (G^{ii} \overline G^{jj} - 2 H^{ii} \overline H^{jj})\;.
\end{multline}
Let us notice that due to the $U(f)$ constraint, the term in $c_2$ is only present when
\begin{equation}
G^{ij} \overline G^{qq} + 2 H^{pp} \overline H^{ij} ~=~ G^{qq} \overline G^{ij} + 2 H^{ij} \overline H^{qq}\;,
\end{equation}
which is indeed the case due to hermiticity.

\subsection{The renormalization factors}
Let us now try to reabsorb this counterterm into the starting action \eqref{CGZ}. We shall split this analysis into three parts, according to
\begin{equation}
\Sigma^c_A + \Sigma^c_{P - H} = \Sigma^c_I + \Sigma^c_{II} + \Sigma^c_{III}\;,
\end{equation}
whereby
\begin{eqnarray}
\Sigma^c_I &=&  \int \d^4x\;  c_1  (G^{ij} \overline G^{ij} - 2 H^{ij} \overline H^{ij}) + c_{2} (G^{ii} \overline G^{jj} - 2 H^{ii} \overline H^{jj})\;, \nonumber\\
\Sigma^c_{II} &=&  \int \d^4x\; b_1 [ (Q+W) A_\mu^a A_\mu^a +2 (P+V) \p_\mu c^a A_\mu^a] +   \frac{a_{2}}{2}\tau A_{\mu }^{a}A_{\mu }^{a} + \left( a_{2}-a_{1}\right) \eta A_{\mu }^{a}\partial _{\mu }c^{a} \;,\nonumber\\
\Sigma^c_{III} &=&  \int \d^4x\;  b_3 Q Q + b_4 Q W + b_6 WW  + b_7 Q \tau + b_8 W \tau + \frac{a_{3}}{2}\zeta \tau ^{2}\;,
\end{eqnarray}
are the three parts which we shall try to absorb separately.\\
\\
Firstly, we start with the vacuum counterterm connected to the arbitrary parameters $c_1$ and $c_2$. If we redefine $c_1$ and $c_2$, we can write
\begin{equation}
\Sigma^c_I =     \int \d^4x\; c_1 \kappa (G^{ij} \overline G^{ij} - 2 H^{ij} \overline H^{ij}) + c_{2}  \lambda (G^{ii} \overline G^{jj} - 2 H^{ii} \overline H^{jj})\;,
\end{equation}
and if we define
\begin{align}
\overline H^{ij}_0 &= Z_{\overline H }  \overline H^{ij}\;,  &  H^{ij}_0 &= Z_{ H }  H^{ij} \;, & \overline G^{ij}_0 &= Z_{\overline G }  \overline G^{ij} \;, & G^{ij}_0 &= Z_{G }  \overline G^{ij}\;, & \kappa_0 &= Z_\kappa \kappa \;, & \lambda_0 &= Z_\lambda \lambda\;,
\end{align}
we find for the renormalization factors of the new sources and the LCO parameters $\kappa$ and $\lambda$:
\begin{eqnarray}\label{Zfac}
Z_{\overline H} &=& Z_{\overline \varphi}^{-1/2} Z_{\overline \omega}^{-1/2} \;,\nonumber\\
Z_{\overline G } &=& Z_{\overline \varphi}^{-1} \;,\nonumber\\
Z_{ H } &=& Z_{ \varphi}^{-1/2} Z_{ \omega}^{-1/2}\;, \nonumber\\
Z_{G } &=& Z_{ \varphi}^{-1}\;,\nonumber\\
Z_{\kappa} &=& (1+ c_1) Z_{\overline G}^{-1} Z_{G}^{-1} = (1+ c_1) Z_{\overline H}^{-1} Z_{H}^{-1} \;,\nonumber\\
Z_{\lambda} &=& (1+ c_2) Z_{\overline G}^{-1} Z_{G }^{-1} = (1+ c_2) Z_{\overline H }^{-1} Z_{H}^{-1}\;,
\end{eqnarray}
and thus the part $\Sigma^c_I$ can absorbed in the starting action.\\
\\
Secondly, let us focus on $\Sigma^c_{II}$
\begin{equation}
\Sigma^c_{II} =  \int \d^4x\;  b_1 [ (Q+W) A_\mu^a A_\mu^a +2 (P+V) \p_\mu c^a A_\mu^a] +   \frac{a_{2}}{2}\tau A_{\mu }^{a}A_{\mu }^{a} + \left( a_{2}-a_{1}\right) \eta A_{\mu }^{a}\partial _{\mu }c^{a}\;.
\end{equation}
We propose the following mixing matrix:
\begin{eqnarray}
 \left(
  \begin{array}{c}
     Q_0 \\
     W_0 \\
     \tau_0
  \end{array}
\right) &= & \left(
          \begin{array}{ccc}
            Z_{QQ}&Z_{QW} &Z_{Q\tau} \\
            Z_{WQ}  &Z_{WW} & Z_{W\tau}   \\
           Z_{\tau Q} & Z_{\tau W}& Z_{\tau\tau}          \end{array}
        \right)
        \left(
\begin{array}{c}
    Q\\
    W \\
    \tau
  \end{array}
\right)\,.
\end{eqnarray}
\begin{itemize}
\item From
\begin{equation}
Q_0 \varphi^a_{i,0} \overline \varphi^a_{i,0} =  [Z_{QQ}Q + Z_{QW} W + Z_{Q\tau} \tau] Z_{\varphi} \varphi^a_{i} \overline \varphi^a_{i} ~=~ Q \varphi^a_{i} \overline \varphi^a_{i}\;,
\end{equation}
we find that $Z_{QQ} = Z_{\overline \varphi}^{-1}$, while $Z_{QW} = Z_{Q\tau}=0$.
\item From
\begin{equation}
W_0 \overline \omega^a_{i,0} \omega^a_{i,0} = [Z_{WQ}Q + Z_{WW} W + Z_{W\tau} \tau] Z_{\varphi}\overline \omega^a_{i} \omega^a_{i} =W \varphi^a_{i} \overline \varphi^a_{i}\;,
\end{equation}
we find that $Z_{WW} = Z_{\overline \varphi}^{-1}$, while $Z_{WQ} = Z_{W\tau}=0$.

\item Finally, from
\begin{multline}
\frac{1}{2}\tau_0 A^a_{\mu,0} A^a_{\mu,0} = \frac{1}{2} [Z_{\tau Q} Q + Z_{\tau W} W + Z_{ \tau \tau} \tau] Z_A  A^a_{\mu} A^a_{\mu}  \\=  \frac{1}{2} \left( 1 +  a_2 \right) \tau A_{\mu}^{a}A_{\mu }^{a} + b_1 Q  A_{\mu }^{a}A_{\mu }^{a} + b_1 W  A_{\mu }^{a}A_{\mu }^{a} \;,
\end{multline}
we obtain $Z_{\tau \tau} = Z_\tau= \left( 1 + a_2 \right)Z_A^{-1}$, and $Z_{\tau Q} = Z_{\tau W} = 2 b_1$.
\end{itemize}
In summary, we find the following matrix
\begin{eqnarray}\label{mixing}
 \left(
  \begin{array}{c}
     Q_0 \\
     W_0 \\
     \tau_0
  \end{array}
\right) &= & \left(
          \begin{array}{ccc}
            Z_{ \varphi}^{-1}& 0 &0 \\
            0  & Z_{\varphi}^{-1} & 0   \\
           Z_{\tau W} & Z_{\tau W}& Z_{\tau\tau}          \end{array}
        \right)
        \left(
\begin{array}{c}
    Q\\
    W \\
    \tau
  \end{array}
\right)\,.
\end{eqnarray}
Now that we have the mixing matrix at our disposal, we can pass to the corresponding bare operators by taking the inverse of this matrix,
\begin{eqnarray}
 \left(
  \begin{array}{c}
     Q \\
     W \\
     \tau
  \end{array}
\right) &= & \left(
          \begin{array}{ccc}
            Z_{\varphi}& 0 &0 \\
            0  & Z_{ \varphi} & 0   \\
           - \frac{ Z_{\tau W} Z_{\varphi} }{ Z_{\tau\tau}} & - \frac{ Z_{\tau W} Z_{ \varphi} }{ Z_{\tau\tau}} & \frac{1}{Z_{\tau\tau}}          \end{array}
        \right)
        \left(
\begin{array}{c}
    Q_0\\
    W_0 \\
    \tau_0
  \end{array}
\right)\,.
\end{eqnarray}
Subsequently, we can derive the corresponding mixing matrix for the operators, since insertions of an operator correspond to derivatives w.r.t.~to the appropriate source of the generating functional $Z^c(Q,W,\tau)$. In particular,
\begin{eqnarray}
\frac{1}{2}      A^2_0 &=& \left. \frac{\delta Z^c(Q,W,\tau)}{\delta \tau_0} \right|_{\tau_0 = 0} \nonumber\\
    &=& \frac{\delta Q}{\delta \tau_0}\frac{\delta Z^c(Q,W,\tau) }{\delta Q}+\frac{\delta W}{\delta \tau_0}\frac{\delta Z^c(Q,W,\tau)}{\delta W } + \frac{\delta \tau}{\delta \tau_0}\frac{\delta Z^c(Q,W,\tau)}{\delta \tau} \nonumber\\
    \Rightarrow  A^2_0 &=&   \frac{1}{Z_{\tau\tau}}   A^2 \;,
\end{eqnarray}
and similarly for $ \overline \varphi^a_{i,0} \varphi^a_{i,0}$ and $\overline \omega^a_{i,0} \omega^a_{i,0} $. We thus need to take the transpose of the previous matrix,
\begin{eqnarray}
 \left(
  \begin{array}{c}
      \overline \varphi^a_{i,0} \varphi^a_{i,0} \\
     \overline \omega^a_{i,0} \omega^a_{i,0}  \\
     A^2_0
  \end{array}
\right) &= & \left(
          \begin{array}{ccc}
            Z_{ \varphi}& 0                     & - \frac{ Z_{\tau W} Z_{ \varphi} }{ Z_{\tau\tau}} \\
            0                    & Z_{ \varphi} & - \frac{ Z_{\tau W} Z_{ \varphi} }{ Z_{\tau\tau}}   \\
            0                     &  0                   &   \frac{1}{Z_{\tau\tau}}          \end{array}
        \right)
        \left(
\begin{array}{c}
     \overline \varphi^a_{i} \varphi^a_{i} \\
     \overline \omega^a_{i} \omega^a_{i} \\
    A^2
  \end{array}
\right)\,.
\end{eqnarray}
We can make some observations from this matrix. Firstly, we find that $A^2_0$ does not contain the operators $ \overline \varphi^a_{i} \varphi^a_{i} $ and $ \overline \omega^a_{i} \omega^a_{i}$. This is already a first check on our results as without these latter two  operators the GZ action including $A^2$ is renormalizable,   as we have shown already in the appendix \ref{appendixB}. Secondly, we observe that
\begin{eqnarray}
\overline \varphi^a_{i,0} \varphi^a_{i,0} - \overline \omega^a_{i,0} \omega^a_{i,0}  &=& Z_\varphi (  \overline \varphi^a_{i} \varphi^a_{i} - \overline \omega^a_{i} \omega^a_{i})\;,
\end{eqnarray}
meaning that the mixing with $A^2$ disappears again when recombining the two operators in a certain way. In fact, this is the operator $(  \overline \varphi^a_{i} \varphi^a_{i} - \overline \omega^a_{i} \omega^a_{i})$ which we have investigated using the RGZ action \cite{Dudal:2008sp} and no mixing with $A^2$ appears for this operator.  \\
\\
We can do a completely analogous reasoning for the part in $\p_\mu c^a A^a_\mu$. We first set $V + P = X$. We propose
\begin{eqnarray}
 \left(
  \begin{array}{c}
     X_0 \\
     \eta_0
  \end{array}
\right) &= & \left(
          \begin{array}{ccc}
            Z_{XX}& Z_{X\eta} \\
           Z_{\eta X} &  Z_{\eta\eta}          \end{array}
        \right)
        \left(
\begin{array}{c}
    X\\
    \eta
  \end{array}
\right)\,.
\end{eqnarray}
\begin{itemize}
\item From
\begin{eqnarray*}
- (X_0)  [\overline \varphi^a_{i,0}  \omega^a_{i,0} ]  &=&  - [ Z_{XX} X + Z_{X \eta}  \eta]  Z_{\overline \varphi}^{1/2} Z_{\omega}^{1/2} \overline \varphi^a_{i} \omega^a_{i} ~=~ - X [\overline \varphi^a_{i}  \omega^a_{i} ]
\end{eqnarray*}
we find that $Z_{XX} = Z_{\overline \varphi}^{-1/2} Z_{\omega}^{-1/2} $, while $Z_{X\eta} =0$.

\item Also, from
\begin{equation*}
\eta_0 A^a_{\mu,0} \p_{\mu} c^a_0 =  [Z_{\eta X} X  + Z_{ \eta \eta} \eta]  Z_A^{1/2} Z_c^{1/2}  A^a_{\mu} \p_{\mu} c^a =  \left( 1 +  a_2-a_1 \right) \eta A^a_{\mu} \p_{\mu} c^a +  2 b_1 X A^a_{\mu} \p_{\mu} c^a \;,
\end{equation*}
we obtain $Z_{\eta \eta} = Z_\eta= \left( 1 +  a_2-a_1 \right) Z_A^{-1/2} Z_c^{-1/2} $, and $Z_{\eta X} = 2 b_1$.
\end{itemize}
Therefore, we find that
\begin{eqnarray}
 \left(
  \begin{array}{c}
      \overline \varphi^a_{i,0} \omega^a_{i,0} \\
     A_{\mu,0} \p_\mu c_0
  \end{array}
\right) &= & \left(
          \begin{array}{ccc}
           Z_A^{1/2} Z_c^{1/2}          & -2 b_1                       \\
           0                     &  Z_{\eta}^{-1}
                     \end{array}
        \right)
        \left(
\begin{array}{c}
     \overline \varphi^a_{i} \omega^a_{i} \\
     A_{\mu} \p_\mu c
  \end{array}
\right)\,.
\end{eqnarray}
Again, we find  that  $A_{\mu,0} \p_\mu c_0$ does not contain $\overline \varphi^a_{i,0} \omega^a_{i,0} $, which is necessary as the GZ action with the inclusion of $A^2$ is renormalizable. We also see that, when setting $V = - P$, $X =0$, the mixing with $A^2$ disappears again.\\
\\
Thirdly, the vacuum term $\Sigma_{III}^c$ has the following form
\begin{eqnarray}
 b_3 Q Q + b_4 Q W + b_6 W W  + b_7 Q \tau + b_8 W \tau + \frac{a_{3}}{2}\zeta \tau ^{2}\;,
\end{eqnarray}
we know that setting $Q = -W$ has to return the vacuum term from the RGZ action $\sim a_4 Q \tau + \frac{a_{3}}{2}\zeta \tau ^{2}$. Therefore, we may set
\begin{eqnarray}
b_3 -b_4 + b_6 = 0 \;.
\end{eqnarray}
In this case, the vacuum term reduces to
\begin{eqnarray}
- c_1 \alpha (Q Q +  Q W) - c_2 \beta( QW + W W)  - c_3 \chi Q \tau - c_4  \delta W \tau  + \frac{a_{3}}{2}\zeta \tau ^{2}\;,
\end{eqnarray}
where we have extracted $\alpha$, $\beta$, $\chi$ and $\delta$ and some minus signs for convenience. If we allow mixing between the different parameters,
\begin{eqnarray}
 \left(
  \begin{array}{c}
      \alpha_0 \\
      \beta_0\\
      \chi_0\\
      \delta_0 \\
      \zeta_0
  \end{array}
\right) &= & \left(
          \begin{array}{ccccc}
           Z_{\alpha\alpha}          & Z_{\alpha\beta}  &  Z_{\alpha\chi}  & Z_{\alpha\delta}  & Z_{\alpha \zeta}         \\
           Z_{\beta\alpha}           & Z_{\beta\beta}   &  Z_{\beta\chi}   & Z_{\beta\delta}   & Z_{\beta \zeta}          \\
           Z_{\chi\alpha}          & Z_{\chi\beta}  &  Z_{\chi\chi}  & Z_{\chi\delta}  & Z_{\chi \zeta}              \\
           Z_{\delta\alpha}          & Z_{\delta\beta}  &  Z_{\delta\chi}  & Z_{\delta\delta}  & Z_{\delta \zeta}               \\
           Z_{\zeta\alpha}           & Z_{\zeta\beta}   &  Z_{\zeta\chi}   & Z_{\zeta\delta}   & Z_{\zeta \zeta}
                     \end{array}
        \right)
        \left(
\begin{array}{c}
 \alpha \\
      \beta \\
      \chi \\
      \delta \\
      \zeta
  \end{array}
\right)\,.
\end{eqnarray}
when absorbing the counterterm, we find for the mixing matrix of the LCO parameters
\begin{equation} \left(
          \begin{array}{ccccc}
           Z_{\alpha\alpha}          & Z_{\alpha\beta}  &  Z_{\alpha\chi}  & Z_{\alpha\delta}  & Z_{\alpha \zeta}         \\
           Z_{\beta\alpha}           & Z_{\beta\beta}   &  Z_{\beta\chi}   & Z_{\beta\delta}   & Z_{\beta \zeta}          \\
           Z_{\chi\alpha}          & Z_{\chi\beta}  &  Z_{\chi\chi}  & Z_{\chi\delta}  & Z_{\chi \zeta}              \\
           Z_{\delta\alpha}          & Z_{\delta\beta}  &  Z_{\delta\chi}  & Z_{\delta\delta}  & Z_{\delta \zeta}               \\
           Z_{\zeta\alpha}           & Z_{\zeta\beta}   &  Z_{\zeta\chi}   & Z_{\zeta\delta}   & Z_{\zeta \zeta}
                     \end{array}
        \right) =
\left(
          \begin{array}{ccccc}
           \frac{1+c_1}{Z_{QQ}^2}        & 0 &  - \frac{Z_{\chi \chi} Z_{\tau W}}{Z_{QQ}} & 0   & \frac{Z_{\tau W}^2 Z_{\zeta \zeta}}{2Z_{QQ}^2}     \\
           0          &\frac{1+c_2}{Z_{QQ}^2}  &  0  & - \frac{Z_{\delta \delta} Z_{\tau W}}{Z_{QQ}}  & \frac{Z_{\tau W}^2 Z_{\zeta \zeta}}{2Z_{QQ}^2}    \\
           0          & 0 &  \frac{1+c_3}{Z_{QQ}Z_{\tau\tau}}  & 0    &  -\frac{ Z_{\tau W} Z_{\zeta \zeta}}{Z_{QQ} }   \\
           0          & 0 &  0  &  \frac{1+c_4}{Z_{QQ}Z_{\tau\tau}}   &  -\frac{ Z_{\tau W} Z_{\zeta \zeta}}{Z_{QQ} }     \\
           0&0&0&0& \frac{1-a_3}{Z_{\tau\tau}^2}
                     \end{array}
        \right)\;.
\end{equation}
In summary, we have proven the action \eqref{actionCGZ} to be renormalizable.

\section{List of propagators}
We give  here the list of propagators which can be calculated from the GZ action \eqref{GZaction}:
\begin{eqnarray}\label{summarypropGZ}
\Braket{\widetilde{ \overline \omega}^{ab}_\mu(k) \widetilde \omega^{cd}_\nu(p)} &=& \delta^{ac}\delta^{bd} \delta^{\mu\nu} \frac{-1}{p^2 } \delta(p+k) (2\pi)^4 \;, \nonumber\\
\Braket{ \widetilde{ \overline c}^a(k) \widetilde c^b(p) } &=& \delta^{ab} \frac{1}{p^2} \delta(p+k) (2\pi)^4 \;,\nonumber\\
\Braket{ \widetilde A_\mu^a (p)  \widetilde A_\nu^b (k)}&=&   \frac{ p^2 }{p^4  + \lambda^4}  P_{\mu \nu} \delta^{ab} \delta(k+ p) (2\pi)^4\;, \nonumber\\
\Braket{\widetilde A_\mu^a(p) \widetilde b^b(k)} &=& - \ii \frac{p_\mu}{ p^2} \delta^{ab} \delta(p+k) (2\pi)^4 \;, \nonumber\\
\Braket{b^a(p) b^b(k)} &=&   \delta^{ab} \frac{ \lambda^4}{p^4 }\delta(p+k)(2\pi)^4 \;, \nonumber\\
\Braket{ \widetilde A^a_{\mu}(p) \widetilde{ \varphi}^{bc}_{\nu}(k)} &=& \Braket{ \widetilde A^a_{\mu}(p) \widetilde{\overline \varphi}^{bc}_{\nu}(k)} ~=~ f^{abc}  \frac{- g \gamma^2}{p^4   + \lambda^4}  P_{\mu \nu}(p) (2\pi)^4 \delta(p+k) \;, \nonumber\\
 \Braket{ \widetilde b^a(p) \widetilde{ \varphi}^{bc}_{\nu}(k)} &=&  \Braket{ \widetilde b^a(p) \widetilde{ \overline \varphi}^{bc}_{\nu}(k)} ~=~  f^{abc} \ii p_\nu \frac{-  g \gamma^2}{ p^4}  (2\pi)^4 \delta(p+k) \;, \nonumber\\
  \Braket{\widetilde{ \varphi}^{ab}_{\mu}(p) \widetilde{ \overline \varphi }^{cd}_{\nu}(k)} &=& \left(  f^{abr} f^{cdr} P_{\mu \nu} \frac{ g^2 \gamma^4}{ p^2( p^4  + 2 g^2 N \gamma^4 )} +  \frac{-1}{ p^2 } \delta^{ac} \delta^{bd} \delta_{\mu \nu}  \right)    (2\pi)^4 \delta(p+k)  \;,\nonumber\\
   \Braket{\widetilde{ \varphi}^{ab}_{\mu}(p) \widetilde{ \varphi }^{cd}_{\nu}(k)} &=&  \Braket{\widetilde{\overline \varphi}^{ab}_{\mu}(p) \widetilde{\overline \varphi }^{cd}_{\nu}(k)} ~=~ f^{abr} f^{cdr} P_{\mu \nu} \frac{ g^2 \gamma^4}{ p^2( p^4  + 2 g^2 N \gamma^4 )}   (2\pi)^4 \delta(p+k) \;.
\end{eqnarray}
with
\begin{equation}
P_{\mu\nu} = \left( \delta_{\mu\nu} - \frac{p_\mu p_\nu}{p^2} \right)  \label{proj} \;.
\end{equation}

\section{Details of the calculation of the effective action for the further refined GZ action\label{detailsvery}}
\subsection{Determination of the LCO parameters $\delta \zeta$, $\delta \alpha$, $\delta \chi$ and $\delta \varrho$ \label{appsection4.2}}
We shall start from expression \eqref{startxx}, determine the quadratic part, and integrate out all the fields. The quadratic action is given by
\begin{multline}\label{start2}
\Sigma_\CGZ^{\quadr} = \int \d^d x  \left[ A_\mu^a \delta^{ab} \left( - \delta_{\mu\nu} \p^2 + \left( 1 - \frac{1}{\alpha}\right) \p_\mu \p_\nu \right)A_\nu^b  + \overline \varphi \p^2 \varphi - \gamma^2 g f_{abc} A_\mu^a (\varphi^{bc}_\mu + \overline \varphi^{bc}_\mu) \right]\nonumber\\
\\+ \int \d^4 x  \left[ Q  \overline \varphi^a_i \varphi^a_i   + \frac{1}{2} \tau A_{\mu }^{a}A_{\mu }^{a}  - \frac{1}{2}\zeta \tau ^{2}  - \alpha   Q Q    - \chi Q \tau\right]  + \int \d^4 x \left[ \frac{1}{2} \overline G \overline \varphi^a_i \overline \varphi^a_i + \frac{1}{2} G \varphi^a_i \varphi^a_i  + \varrho G \overline G \right]\;,
\end{multline}
whereby we have immediately integrated out the ghost fields, $c, \overline c, \omega, \overline \omega$, as they  only appear trivially. We have also already integrated out the $b$-field whereby $\alpha$ is formally equal to zero.\\
\\
As a first step, we integrate out the $\varphi$ and $\overline \varphi$ fields. For this, we shall split $\varphi$, $\overline \varphi$, $G$ and $\overline G$ into real  and imaginary components:
\begin{align}
\overline \varphi^a_i &= U^a_i + \ii V^a_i \;,   & \varphi^a_i &= U^a_i - \ii V^a_i  \;,\nonumber\\
\overline G &= X + \ii Y  \;,  & G &= X- \ii Y \;,
\end{align}
so  that the part depending on $\varphi$ and $\overline \varphi$ in expression \eqref{start2} becomes
\begin{eqnarray*}
&&\int \d^d x \left( U^a_i\p^2 U^a_i + V^a_i \p^2 V^a_i  - 2 \gamma^2 g f_{abc} A_\mu^a U^{bc}_\mu + Q U^2 + Q V^2 + X U^2 - X V^2 \right. \nonumber\\
 && \left.- 2 Y U^a_iV^a_i + \varrho X^2 + \varrho Y^2 \right) \nonumber\\
&=& \int \d^d x \left( \frac{1}{2} \begin{bmatrix}
     U_\mu^{ab} &      V_\mu^{ab}
  \end{bmatrix}   \begin{bmatrix}
    2 (\p^2 + Q + X) &      -2 Y \\
    -2Y &  2 (\p^2 + Q - X)
  \end{bmatrix} \begin{bmatrix}
     U_\mu^{ab} \\     V_\mu^{ab}
  \end{bmatrix}   - 2 \gamma^2 g f_{abc} A_\mu^a U^{bc}_\mu \right)\;.
\end{eqnarray*}
Therefore, applying Gaussian integration, we find for the integration over $\varphi$ and $\overline \varphi$
\begin{equation} \label{uitinto}
\int [\d \varphi] [\d \overline \varphi] \exp [- \Sigma_\CGZ^{\quadr} ] = \exp \left[ \frac{1}{2} \lambda^4 A_\mu^k \left( \frac{\p^2 + Q - X}{\p^4 + 2 Q \p^2 + Q^2 - X^2 - Y^2}\right) A_\mu^k + \ldots\right] (\det P_{\mu\nu}^{ab,cd})^{-1/2}\;,
\end{equation}
whereby we recall that $\lambda$ is defined as $\lambda^4 =  2 \gamma^4 g^2 N$.  $P_{\mu\nu}^{ab,cd} $ is given by
\begin{equation}
P_{\mu\nu}^{ab,cd} =\delta_{\mu\nu} \delta^{ab} \delta^{cd}\begin{bmatrix}
    2 (\p^2 + Q + X) &      -2 Y \\
    -2Y &  2 (\p^2 + Q - X)
  \end{bmatrix}\;,
\end{equation}
and the $\ldots$ stand for the other terms in $\Sigma_\CGZ^{\quadr}$, see \eqref{start2}, i.e.~terms purely in $A$ and the vacuum terms. The second step is to integrate out the gluon field $A_\mu^a$. Combining the expression \eqref{uitinto} with the terms purely in $A$ from the quadratic action, we obtain,
\begin{multline}
\int [\d A] \e^{ \Bigl[- \frac{1}{2}  A_\mu^a \delta^{ab} \left( - \delta_{\mu\nu} \p^2 + \left( 1 - \frac{1}{\alpha}\right) \p_\mu \p_\nu  -  \lambda^4 \left( \frac{\p^2 + Q - X}{\p^4 + 2 Q \p^2 + Q^2 - X^2 - Y^2}\right) + \tau \delta_{\mu\nu}  \right)A_\nu^b \Bigr]}\\
= \left[ \det \left( - \delta_{\mu\nu} \p^2 + \left( 1 - \frac{1}{\alpha}\right) \p_\mu \p_\nu  -  \lambda^4 \left( \frac{\p^2 + Q - X}{\p^4 + 2 Q \p^2 + Q^2 - X^2 - Y^2}\right) + \tau \delta_{\mu\nu}  \right) \right]^{-1/2}\;.
\end{multline}
Therefore, the total effective action at one loop is given by
\begin{multline}\label{totaleffaction}
\e^{-W(Q, \tau, G, \overline G)}=   (\det P_{\mu\nu}^{ab,cd})^{-1/2} \Bigl[ \det \Bigl( - \delta_{\mu\nu} \p^2 + \left( 1 - \frac{1}{\alpha}\right) \p_\mu \p_\nu \\ -  \lambda^4 \delta_{\mu\nu}\left( \frac{\p^2 + Q - X}{\p^4 + 2 Q \p^2 + Q^2 - X^2 - Y^2}\right) + \tau \delta_{\mu\nu}  \Bigr) \Bigr]^{-1/2} \e^{\left[ - \int \d^4 x \left[ - \frac{1}{2}\zeta \tau ^{2}  - \alpha   Q Q    - \chi Q \tau \frac{1}{2}  + \varrho G \overline G \right]\right] }\;.
\end{multline}
In order to find $\delta \zeta$, $\delta \alpha$, $\delta \chi$ and $\delta \varrho$ at one loop, we need to find the first order infinities of the previous expression. These shall be present in the two determinants which we need to evaluate.\\
\\
Let us start with the first determinant of  $P_{\mu\nu}^{ab,cd}$. In general, we can write
\begin{equation}
(\det P_{\mu\nu}^{ab,cd} )^{-1/2} = \e^{-\frac{1}{2} \Tr \ln P_{\mu\nu}^{ab,cd}} = \e^{-\frac{1}{2} d (N^2-1)^2 \Tr \ln P}\;.
\end{equation}
As we are taking the trace, we know that $\Tr \ln P = \Tr \ln P'$ with $P'$ the diagonalization of $P$. Therefore, after diagonalization, we find
\begin{multline}
(\det P_{\mu\nu}^{ab,cd})^{-1/2} \\= \exp \left[ -\frac{1}{2} d (N^2-1)^2 \Tr  \left( \ln ( -\p^2 - Q + \sqrt{X^2 + Y^2}) + \ln ( -\p^2 - Q - \sqrt{X^2 + Y^2}) \right)   \right]\;.
\end{multline}
Employing the standard formula, \cite{Peskin}
\begin{equation}\label{standard2}
\Tr \ln (-\p^2 + M^2) = - \frac{\Gamma(-d/2) }{(4\pi)^{d/2}} \frac{1}{ (M^2)^{-d/2}}\;,
\end{equation}
we obtain the following infinity
\begin{eqnarray}\label{res1}
(\det P)^{-1/2} &=& \exp \left[\frac{1}{\epsilon}\frac{ (N^2 - 1)^2}{4 \pi^2} \left[ Q^2 + X^2 + Y^2\right] + c_1\right]\;,
\end{eqnarray}
whereby $c_1$ is a constant term. \\
\\
The second determinant requires a bit more effort to  be evaluated. Let us call the corresponding matrix $K$. We thus calculate
\begin{equation}
(\det K_{\mu\nu}^{ab})^{-1/2} = \e^{-\frac{1}{2} (N^2-1) \Tr \ln K_{\mu\nu} }\;,
\end{equation}
Therefore, we need to determine
\begin{multline}\label{terms}
\Tr \ln K_{\mu\nu} = \Tr \ln \left(   \delta_{\mu\nu}\left( -\p^2  -  \lambda^4 \left( \frac{\p^2 + Q - X}{\p^4 + 2 Q \p^2 + Q^2 - X^2 - Y^2}\right) + \tau \right)\right) \\ + \Tr \ln \left( \delta_{\mu\nu} + \frac{1}{\left( -\p^2  -  \lambda^4 \left( \frac{\p^2 + Q - X}{\p^4 + 2 Q \p^2 + Q^2 - X^2 - Y^2}\right) + \tau  \right)}  \left( 1 - \frac{1}{\alpha}\right) \p_\mu \p_\nu  \right)\;.
\end{multline}
For the first term, we can easily take the trace over the Lorentz indices, while for the second term, we need to use $\ln (1 + x) = x - \frac{x^2}{2} + \ldots$, then take the trace of the diagonal elements of the second term, and again employ $x - \frac{x^2}{2} + \ldots = \ln (1 + x)$. After these operations, we obtain
\begin{multline*}
\Tr \ln K_{\mu\nu} = d \Tr \ln \left(  \left( -\p^2  -  \lambda^4 \left( \frac{\p^2 + Q - X}{\p^4 + 2 Q \p^2 + Q^2 - X^2 - Y^2}\right) + \tau \right)\right) \\ + \Tr \ln \left( 1 + \frac{1}{\left( -\p^2  -  \lambda^4 \left( \frac{\p^2 + Q - X}{\p^4 + 2 Q \p^2 + Q^2 - X^2 - Y^2}\right) + \tau  \right)}  \left( 1 - \frac{1}{\alpha}\right) \p^2  \right)\;,
\end{multline*}
which can be written as
\begin{multline*}
\Tr \ln K_{\mu\nu} = (d-1) \Tr \ln \left(  \left( -\p^2  -  \lambda^4 \left( \frac{\p^2 + Q - X}{\p^4 + 2 Q \p^2 + Q^2 - X^2 - Y^2}\right) + \tau \right)\right) \\ + \Tr \ln \left( \left( -\p^2  -  \lambda^4 \left( \frac{\p^2 + Q - X}{\p^4 + 2 Q \p^2 + Q^2 - X^2 - Y^2}\right) + \tau  \right) +   \left( 1 - \frac{1}{\alpha}\right) \p^2  \right)\;.
\end{multline*}
The first term of this expression can be written as\footnote{We shall replace $-\p^2$ by $p^2$ from now on and work in momentum space.}
\begin{align}
& (d-1) \Bigl[ \Tr \ln \bigl(  p^6+(\tau -2 Q) p^4+\left(\lambda ^4+Q^2-X^2-Y^2-2 Q \tau \right) p^2-Q \lambda ^4+X \lambda ^4+Q^2 \tau \nonumber\\
 &-X^2 \tau -Y^2 \tau  \bigr)  -\Tr \ln \left(  p^4-2 Q p^2+Q^2-X^2-Y^2 \right) \Bigr]\nonumber\\
=& (d-1) \left( \Tr \ln (p^2 - x_1) + \Tr \ln (p^2 - x_2) + \Tr \ln (p^2 - x_3) - \Tr \ln (p^2 - x_4) - \Tr \ln (p^2 - x_5)  \right)\;,
\end{align}
whereby $x_1$, $x_2$ and $x_3$ are the solutions of the equation $x^3+(\tau -2 Q) x^2+\bigl(\lambda ^4+Q^2-X^2-Y^2-2 Q \tau \bigr) x -Q \lambda ^4+X \lambda ^4+Q^2 \tau -X^2 \tau -Y^2 \tau =0$
and $x_4$ and $x_5$ of the equation $ x^2 -2 Q x +Q^2-X^2-Y^2 =0$. After determining $x_1, \ldots, x_5$, we can apply the standard formula \eqref{standard2} again, so we ultimately find for the first term
\begin{equation}
- \frac{3}{16 \pi^2} \frac{1}{\epsilon} \left( \tau^2 - 2 \lambda^4 \right) + c_2\;,
\end{equation}
with $c_2$ a constant, which is not  of our current interest. For the second term of \eqref{terms}, we can perform an analogous analysis, whereby we find that this term is proportional to $\alpha$ and therefore does not contribute to the determinant as $\alpha \to 0$. Therefore, the second determinant ultimately gives:
\begin{equation}\label{res2}
(\det K_{\mu\nu}^{ab})^{-1/2} = \exp\left[  (N^2-1)  \frac{3}{32 \pi^2} \frac{1}{\epsilon} \left( \tau^2 - 2 \lambda^4 \right) + c_2 \right] \;.
\end{equation}
We can now combine both results \eqref{res1} and \eqref{res2} to find
\begin{equation}
W(Q, \tau, G, \overline G) = - \frac{(N^2 -1)}{4 \pi^2} \frac{1}{\epsilon} \left( \frac{3}{8}\tau^2 + (N^2-1)( Q^2 + G \overline G) - \frac{3}{4} \lambda^4 \right) + c \;,
\end{equation}
with $c$ a constant term. Therefore, at one loop we obtain
\begin{eqnarray}\label{deltas}
\delta \zeta &=& -\frac{1}{\epsilon} \frac{3}{16 \pi^2} (N^2-1) \;,\nonumber\\
\delta \alpha &=& -\frac{1}{\epsilon} \frac{1}{4 \pi^2} (N^2-1)^2 \;,\nonumber\\
\delta \chi &=& 0 \;, \nonumber\\
\delta \varrho &=& \frac{1}{\epsilon} \frac{1}{4 \pi^2} (N^2-1)^2\;.
\end{eqnarray}

\subsection{Calculation of the effective action}
We can now proceed in a very similar fashion as in section \ref{appsection4.2}. We can split the one loop effective potential in a few parts. A first part, $\Gamma^{(1)}_a$, is the equivalent of $(\det P)^{-1/2}$ in expression \eqref{totaleffaction}
\begin{multline}
\Gamma^{(1)}_a = (N^2 -1)^2 \Biggl[ -\frac{1}{\epsilon} \frac{1}{4\pi^2} (M^4 + \rho \rho^\dagger) + \frac{1}{16 \pi^2} \left( (M^2 - \sqrt{\rho \rho^\dagger})^2 \ln\frac{M^2 - \sqrt{\rho \rho^\dagger}}{\overline \mu^2} \right. \\
\left. + (M^2 + \sqrt{\rho \rho^\dagger})^2 \ln \frac{M^2 + \sqrt{\rho \rho^\dagger}}{\overline \mu^2} - 2 (M^2 + \rho \rho^\dagger) \right) \Biggr]\;.
\end{multline}
The second part, the equivalent of $(\det K)^{-1/2}$, is given by
\begin{multline}
\Gamma^{(1)}_b = \frac{3(N^2 -1)}{64\pi^2} \Biggl[ -\frac{2}{\epsilon}  (m^4 -2 \lambda^4) - \frac{5}{6} (m^4 -2 \lambda^4) + y_1^2 \ln \frac{(-y_1)}{\overline \mu} + y_2^2 \ln \frac{(-y_2)}{\overline \mu} + y_3^2 \ln \frac{(-y_3)}{\overline \mu} \\- y_4^2 \ln \frac{(-y_4)}{\overline \mu} - y_5^2 \ln \frac{(-y_5)}{\overline \mu} \Biggr]\;,
\end{multline}
whereby $y_1$, $y_2$ and $y_3$ are the solutions of the equation $y^3+(m^2 +2 M^2) y^2 +\bigl(\lambda^4+ M^4- \rho \rho^\dagger +2 M^2 m^2 \bigr) y + M^2 \lambda^4 + 1/2 ( \rho + \rho^\dagger) \lambda^4 + M^4 m^2  - m^2 \rho \rho^\dagger  =0$
and $y_4$ and $y_5$ of the equation $ y^2 + 2 M^2 y +M^4 -\rho \rho^\dagger =0$. \\
The third part is the constant term of the GZ action,
\begin{equation}
\Gamma^{(1)}_c = -d \gamma^4_0 (N^2 - 1)\;.
\end{equation}
From equation \eqref{Zgamma}, we can calculate that\footnote{For the explicit loop calculations of the $Z$-factors, we refer to \cite{Gracey:2002yt}.}
\begin{eqnarray}
\gamma_0^4 &=& Z_{\gamma^2}^2 \gamma^4\;,  \qquad \text{with} \qquad Z_{\gamma^2}^2 = 1+ \frac{3}{2} \frac{g^2N}{16 \pi^2} \frac{1}{\epsilon}\;,
\end{eqnarray}
so we find
\begin{eqnarray}
\Gamma^{(1)}_c&=& -d(N^2 - 1)\gamma^4_0 = -4 (N^2 - 1)\gamma^4 - 4 \frac{3}{2} (N^2 - 1)\frac{g^2 N}{16\pi^2}\frac{1}{\epsilon} \gamma^4 + \frac{3}{2} \frac{g^2N}{16 \pi^2}\gamma^4 (N^2-1) \nonumber\\
&=& -2 (N^2 - 1) \frac{\lambda^4}{N g^2} - 6 (N^2 - 1)\frac{\lambda^4}{32\pi^2}\frac{1}{\epsilon}  + \frac{3}{2} \frac{\lambda^4}{32 \pi^2} (N^2-1)  \;.
\end{eqnarray}
The fourth part requires some calculation. We firstly find
\begin{equation}
 \frac{1}{ 4 Z_\varrho Z_G^2  \varrho }\left( \frac{\sigma_3^2}{g^2} + \frac{\sigma_4^2}{g^2} \right) = \frac{1}{2} \frac{48 (N^2-1)^2}{53 N} \left( 1 - \frac{53}{6} \frac{1}{\epsilon} \frac{N g^2}{16 \pi^2}  - N g^2 \frac{53}{24} \frac{\varrho_1}{(N^2 - 1)^2} \right) \frac{\rho \rho^\dagger}{g^2}\;,
\end{equation}
and secondly
\begin{multline}
\frac{ \alpha' }{4 \alpha' \zeta' - \chi^{\prime 2}}  \frac{\sigma_1^2}{g^2} + \frac{ \zeta' }{ 4 \alpha' \zeta' - \chi^{\prime 2}}  \frac{\sigma_2^2}{g^2}   -  \frac{ \chi' }{4 \alpha' \zeta' - \chi^{\prime 2}}  \frac{\sigma_1 \sigma_2}{g^2} \\= \frac{\zeta_0 m^4}{2 g^2}+ \frac{ \alpha_0 M^4}{g^2} + \frac{1}{\epsilon}\left( \frac{13 N \zeta_0 m^4}{96 \pi ^2}+\frac{M^4 (N^2-1)^2}{4 \pi ^2} \right)  -\frac{\zeta_1 m^4}{2}- M^4 \alpha_1 + M^2 m^2 \chi_1\;,
\end{multline}
so that
\begin{multline}
\Gamma^{(1)}_d = \frac{1}{2} \frac{48 (N^2-1)^2}{53 N} \left( 1 - \frac{53}{6} \frac{1}{\epsilon} \frac{N g^2}{16 \pi^2}  - N g^2 \frac{53}{24} \frac{\varrho_1}{(N^2 - 1)^2} \right) \frac{\rho \rho^\dagger}{g^2} + \frac{\zeta_0 m^4}{2 g^2}+\frac{ \alpha_0 M^4}{g^2}\\
+ \frac{1}{\epsilon}\left( \frac{13 N \zeta_0 m^4}{96 \pi ^2}+\frac{M^4 (N^2-1)^2}{4 \pi ^2} \right)  -\frac{\zeta_1 m^4}{2}- M^4 \alpha_1 + M^2 m^2 \chi_1\;.
\end{multline}
As a check on our results, we see that all the infinities cancel, so we find
\begin{align}
\Gamma^{(1)} &=  \frac{(N^2 -1)^2}{16 \pi^2} \Bigl[(M^2 - \sqrt{\rho \rho^\dagger})^2 \ln\frac{M^2 - \sqrt{\rho \rho^\dagger}}{\overline \mu^2}  + (M^2 + \sqrt{\rho \rho^\dagger})^2 \ln\frac{M^2 + \sqrt{\rho \rho^\dagger}}{\overline \mu^2} \nonumber\\
& - 2 (M^2 + \rho \rho^\dagger)\Bigr] + \frac{3(N^2 -1)}{64\pi^2} \Bigl[  - \frac{5}{6} (m^4 -2 \lambda^4) + y_1^2 \ln \frac{(-y_1)}{\overline \mu} + y_2^2 \ln \frac{(-y_2)}{\overline \mu} + y_3^2 \ln \frac{(-y_3)}{\overline \mu} \nonumber\\
&- y_4^2 \ln \frac{(-y_4)}{\overline \mu} - y_5^2 \ln \frac{(-y_5)}{\overline \mu} \Bigr]  -2 (N^2 - 1) \frac{\lambda^4}{N g^2}  + \frac{3}{2} \frac{\lambda^4}{32 \pi^2} (N^2-1) \nonumber\\
&+ \frac{1}{2} \frac{48 (N^2-1)^2}{53 N} \left( 1   - N g^2 \frac{53}{24} \frac{\varrho_1}{(N^2 - 1)^2} \right) \frac{\rho \rho^\dagger}{g^2} \nonumber\\
&  +  \frac{9}{13} \frac{N^2-1}{N}\frac{m^4}{2g^2}- \frac{24}{35}\frac{(N^2-1)^2}{N}\frac{M^4}{g^2}  - \frac{161}{52} \frac{N^2-1}{16 \pi^2}\frac{ m^4}{2 }- M^4 \alpha_1  +M^2 m^2 \chi_1\;.
\end{align}

\end{document}